\documentclass[final]{agujournal2019}
\usepackage{url} 
\usepackage{lineno}
\usepackage[inline]{trackchanges} 
\usepackage{soul}
\newcommand{\doc}{paper}

\newcommand{\aposteriori}{\textit{a posteriori}}

\newcommand{\modelerror}{model error}

\newcommand{\method}{MEDIDA}

\newcommand{\stateparnull}{\ensuremath{ {w} }} 

\newcommand{\modelstateparc}{\ensuremath{ \stateparnull(\grid,t) }} 

\newcommand{\statedis}[1]{\ensuremath{ {\stateparnull} }} %
\newcommand{\stateobs}[1]{\ensuremath{ \statedis{}^o_{#1} }} %
\newcommand{\statemodel}[1]{\ensuremath{ \statedis{}^m_{#1} }} %
\newcommand{\grid}{\ensuremath{ {x} }} 

\newcommand{\timepre}{\ensuremath{ t_{i}-\Delta t }} 
\newcommand{\timenow}{\ensuremath{ t_{i} }} 

\newcommand{\minimize}{\ensuremath{ \text{min} }}

\newcommand{\Norm}[1]{\ensuremath{  \left\| #1 \right\| }}

\newcommand{\NormT}[1]{\ensuremath{  \Norm{#1}_2 }}

\newcommand{\transpose}[1]{\ensuremath{ {#1}^{\top} }}

\newcommand{\normL}[1]{\ensuremath{ \mathcal{L}_{#1}} }

\newcommand{\error}{\ensuremath{ \varepsilon }}

\newcommand{\order}[1]{\ensuremath{ \mathcal{O} \left(#1\right) }}  
  
\newcommand{\timestep}{\ensuremath{\Delta t}}

\newcommand{\numsamples}{\ensuremath{n}}

\newcommand{\coeffsymbole}{\ensuremath{{c}}}
\newcommand{\coeff}{\ensuremath{\bm{\coeffsymbole}}}

\newcommand{\coefftrue}{\ensuremath{ {\coeff}_{s} }}
\newcommand{\coeffdiscovered}{\ensuremath{ \coeff^* }}
\newcommand{\coeffmodel}{\ensuremath{ \coeff_m }}

\newcommand{\basis}{\ensuremath{ {\phi} }}

\newcommand{\funmodelnull}{\ensuremath{ f }} 
\newcommand{\funtruthnull}{\ensuremath{ g }} 
\newcommand{\funerrornull}{\ensuremath{ h }}

\newcommand{\funmodel}[1]{\ensuremath{ \funmodelnull\left(#1\right) }} 
\newcommand{\funtruth}[1]{\ensuremath{ \funtruthnull\left(#1\right) }} 
\newcommand{\funerror}[1]{\ensuremath{ \funerrornull\left(#1\right) }}

\newcommand{\errorcoeffmodel}{\ensuremath{ \error_m }}
\newcommand{\errorcoeffdiscovered}{\ensuremath{ \error^*} }

\newcommand{\numlibcol}{\ensuremath{ b }}

\newcommand{\inputvec}{\ensuremath{ \bm{u} }}

\newcommand{\topography}[4]{
	{#1} \mathcal{N}{\left( 
		\left[#2,#3\right]
		,#4\right)}
}
\usepackage[acronym]{glossaries}
\newacronym{medida}{MEDIDA}{Model Error Discovery with Interpretability and Data Assimilation}
\newacronym{da}{DA}{data assimilation}
\newacronym{enkf}{EnKF}{ensemble Kalman filter}
\newacronym{enks}{EnKS}{ensemble Kalman smoother}
\newacronym{pde}{PDE}{partial differential equation}
\newacronym{ode}{ODE}{ordinary differential equation}

\newacronym{rvm}{RVM}{relevance vector machine}
\newacronym{lasso}{LASSO}{least absolute shrinkage and selection operator}
\newacronym{sindy}{SINDy}{sparse identification of nonlinear dynamics}
\newacronym{ml}{ML}{machine learning}
\newacronym{nn}{NN}{deep neural network}
\newacronym{mlp}{MLP}{multi-layer perceptron}
\newacronym{ntk}{NTK}{neural tangent kernel}
\newacronym{dnn}{DNN}{deep neural network}
\newacronym{ann}{ANN}{artificial neural network}
\newacronym{cnn}{CNN}{convolutional neural network}
\newacronym{ks}{KS}{Kuramoto-Sivashinsky}
\newacronym{etdrk4}{ETDRK4}{exponential time differencing fourth-order Runge-Kutta}
\newacronym{gcm}{GCM}{global circulation model}
\newacronym{nwp}{NWP}{numerical weather prediction}
\newacronym{mooam}{MAOOAM}{modular arbitrary-order-ocean-atmosphere model}
\newacronym{qg}{QG}{quasi-geostrophic}
\newacronym{rff}{RFF}{random Fourier feature}
\newacronym{rmse}{RMSE}{root-mean-square error}

\newacronym{ad}{AD}{automatic differentiation}
\newacronym{fd}{FD}{finite difference}

\newacronym{cfl}{CFL}{Courant--Friedrichs--Lewy}

\newacronym{pinn}{PINN}{physics-informed neural network}
\newacronym{xpinn}{XPINN}{extended \gls{pinn}}
\newacronym{lpinn}{LPINN}{Lagrangian physics--informed neural network}

\newacronym{lstm}{LSTM}{Long short--term memory}

\newacronym{rom}{ROM}{reduced order model}
\newacronym{lspg}{LSPG}{least--square Petrov--Galerkin}
\newacronym{npm}{NPM}{Neural Particle Method}
\newacronym{rbf}{RBF}{radial basis function}

\newacronym{ado}{ADO}{alternating direction optimization}

\newacronym{gep}{GEP}{gene expression programming}

\newacronym{tcr}{TCR}{transient climate response}
\newacronym{esm}{ESM}{Earth system model}
\newcommand{\streamfunction}[1]{\ensuremath{\psi}_{#1}}

\newcommand{\QGoneRHS}{\ensuremath{ 
		- J\left(\streamfunction{1}, q_{1}\right) 
		+ \frac{1}{ \tau_{{d}_1}}\left(\streamfunction{1}-\streamfunction{2}\right)  
		- \frac{1}{ \tau_{{d}_2}} \streamfunction{R}
		- \nu \nabla^8 \streamfunction{1}
}}

\newcommand{\QGtwoRHS}{\ensuremath{ 
		- J\left(\streamfunction{2}, q_{2}\right) 
		- \frac{1}{ \tau_{{d}_1}}\left(\streamfunction{1}-\streamfunction{2}\right)  
		+ \frac{1}{ \tau_{{d}_2}} \streamfunction{R}
		-\frac{1}{\tau_{f}}  \nabla^{2} \streamfunction{2} 
		-\nu \nabla^8 \streamfunction{2}
}}

\def\mydepository{\url{www.github.com/rmojgani/MEDIDA_QG}}

\usepackage{amsfonts}
\usepackage{amsmath}
\usepackage{bm}
\justifying 
\usepackage[capitalize]{cleveref}
\crefname{appendix}{}{}

\newcommand{\mycolort}{black}

\newcommand{\revt}[1]{{\color{\mycolort}#1}}

\usepackage{graphicx}
\usepackage{multirow}
\DeclareMathOperator{\sech}{sech}
\usepackage{caption}
\usepackage{subcaption}
\usepackage{booktabs}

\usepackage{pifont}
\def\mycros{\ding{55}} 
\usepackage{makecell}

\draftfalse

\journalname{Journal of Advances in Modeling Earth Systems (JAMES)}

\begin{document}

%
%




\title{
{Interpretable structural model error discovery from sparse assimilation increments using spectral bias-reduced neural networks: A quasi-geostrophic turbulence test case}
}
%
%




\authors{Rambod Mojgani\affil{1}, Ashesh Chattopadhyay\affil{1,2}, Pedram Hassanzadeh\affil{1,3}}

\affiliation{1}{Departments of Mechanical Engineering, Rice University, Houston, Texas}
\affiliation{2}{Department of Applied Mathematics, University of California, Santa Cruz, California}
\affiliation{3}{Departments of Earth, Environmental and Planetary Sciences, Rice University, Houston, Texas}




\correspondingauthor{Rambod Mojgani}{rm99@rice.edu}



\begin{keypoints}
\item Model error discovery with interpretability and data assimilation (MEDIDA) is scaled up to geostrophic turbulence and sparse observations
\item Naive use of neural nets (NNs) as interpolator does not capture small scales due to spectral bias, failing discoveries of closed-form errors
\item Reducing this bias using random Fourier features enables NNs to represent the full range of scales, leading to successful error discoveries
\end{keypoints}

%
%

%
%


\begin{abstract}
Earth system models suffer from various structural and parametric errors in their representation of nonlinear, multi-scale processes, leading to uncertainties in their long-term projections. The effects of many of these errors (particularly those due to fast physics) can be quantified in short-term simulations, e.g., as differences between the predicted and observed states (analysis increments). With the increase in the availability of high-quality observations and simulations, learning nudging from these increments to correct model errors has become an active research area. However, most studies focus on using neural networks, which while powerful, are hard to interpret, are data-hungry, and poorly generalize out-of-distribution.  Here, we show the capabilities of Model Error Discovery with Interpretability and Data Assimilation (MEDIDA), a general, data-efficient framework that uses sparsity-promoting equation-discovery techniques to learn model errors from analysis increments. Using two-layer quasi-geostrophic turbulence as the test case, MEDIDA is shown to successfully discover various linear and nonlinear structural/parametric errors when full observations are available. Discovery from spatially sparse observations is found to require highly accurate interpolation schemes. While NNs have shown success as interpolators in recent studies, here, they are found inadequate due to their inability to accurately represent small scales, a phenomenon known as spectral bias.  We show that a general remedy, adding a random Fourier feature layer to the NN, resolves this issue enabling MEDIDA to successfully discover model errors from sparse observations. These promising results suggest that with further development, MEDIDA could be scaled up to models of the Earth system and real observations.
\end{abstract}

\section*{Plain Language Summary}
{Numerical models are used to predict the Earth system, for example, from daily weather to the next-century climate. These models have been developed and validated against observations over decades, however, they still have shortcomings (errors) in their representations of many complex processes, particularly those that are nonlinear and span many scales in time and space. The rapid improvements in the quality and quantity of observational data from the Earth system and advances in machine learning (ML) algorithms provide an opportunity to try to reduce these errors. However, the challenge is that many ML methods require a lot of training data, and it is also often difficult to explain how they are reducing the error. Here, we show the capabilities of a framework called MEDIDA (Model Error Discovery with Interpretability and Data Assimilation), which uses a class of ML methods that provide closed-form (thus interpretable) equations for what they are learning from the differences between observations and model predictions. We show the success of MEDIDA when applied to a model of atmospheric turbulent circulation. Even when observations are only sparsely (not at every location) available, we show that MEDIDA works accurately once we leverage more recent advances in ML. 
}

\section{Introduction}
\label{sec:introduction}


Numerical solutions of physics-based models are the core of modern weather and climate predictions.
However, climate and \gls{nwp} models suffer from a variety of parametric and structural errors ({\it model errors}, hereafter). These model errors often arise from approximations to forcings and boundary conditions, and more importantly, poor representations of complex, usually small-scale, processes, due to the lack of scientific understandings and/or limitations of computational resources~\cite{Stevens_Science_2013,RN4,MOON_JGRA_2018,Woldemeskel_JGRA_2012,Zadra_2018}. Reducing these model errors through improving the fundamental understanding of these processes and increasing the numerical resolution has been the subject of extensive past and ongoing efforts~\cite{Danforth_GRL_2008,Bonavita2_JAMES_2020,Dunbar_JAMES_2021,Regayre_ACP_2023}. More recently, the rapid rise in the availability of high-quality, frequent observations and algorithmic advances, particularly in \gls{da} and \gls{ml}~\cite{Cheng_IEEE_2023}, can provide another promising direction for reducing such model errors and the resulting biases and uncertainties. 

Many model errors for the Earth system are associated with the ``fast physics''~\cite{Rodwell_QJRMS_2007}, and therefore, they can be observed and quantified in short-time prediction horizons. 
Such model errors can lead to deviations of the predicted states from the \revt{analysis states}; such deviations are sometimes referred to as {\it DA analysis increment}~\cite{Leith_ARFM_1987}. \revt{Assuming the analysis states to be our best estimates of the true states of the underlying system, hereafter we treat them as truth.} In short-time forecasts with \glspl{nwp}, these deviations are corrected using observations and classical \gls{da}, e.g., the predicted states are nudged towards the observations. The analysis increments can be also examined to diagnose the roots of these model errors, potentially, helping with improving the long-term weather forecasts and climate projections as well~\cite{Rodwell_QJRMS_2008,Palmer_GAFD_2011}.

The latter perspective and the recent advances in ML have motivated many recent studies to leverage deviations of short-time predictions from observations and/or high-fidelity simulations to correct biases of \gls{nwp} and climate models to improve long-term predictions and projections. The key underlying assumption of this approach is that correcting the errors arising from the fast physics can correct the model's predictions over long time scales. In this approach, the imperfect (numerical) models are initialized with the true state or its close estimate (\revt{e.g., analysis state, }from observations\revt{,} or high-fidelity simulations), evolved forward in time for a short period, and the difference between the predicted and the true states, the ``analysis increment'', is quantified/parameterized as a representative of the model error.

A number of recent studies have followed this approach and used ML to learn the model errors, showing promising results in various systems, from simple toy models to \gls{nwp} models~\cite{Carrassi_IJBC_2011, Mitchell_QJRMS_2015, Lang_Tellus_2016, watson2019applying, pawar2020long, pathak2020using, Farchi_RMetS_2020, Wattmeyer_GRL_2021, yuval2021use, Chen_JAMES_2022, Bretherton_2022_JAMES, Chen_PhysicaD_2022, Mojgani_Chaos_2022, Gregory_arxiv_2023, clark2022correcting, Arcomano_GRL_2023, Bora_arxiv_2023}.
However, most of these studies have directly learned the model error using deep \glspl{ann}: \glspl{ann} are trained using many pairs of state and analysis increments from a training set. Then, for out-of-sample states (that are from the same distribution as those of the training), the ANN predicts the \revt{ systematic model tendency correction} needed to \revt{nudge} the state \revt{of} \gls{nwp} or climate model, thus improving the trajectory and potentially the simulated statistics. While \glspl{ann} are powerfully expressive, they have a number of major shortcomings: 1) they are difficult to interpret, 2) they do not generalize to out-of-distribution, 3) they are data-hungry, and 4) their predictions fed into numerical models can cause instabilities and unphysical drifts \cite{Guan_JCP_2022, subel2023explaining, Bretherton_2022_JAMES, clark2022correcting, Slater_JAMES_2023, Farchi_JAMES_2023, pahlavanQBO}.
Challenges with interpretability hinder understanding the root cause(s) of the model errors and fixing them. This and (2) together significantly limit the applicability of such ANN-based approaches to improve climate change projections. (3) and (4) also pose major practical and operational limitations. The latter is particularly challenging as rigorous frameworks for quantifying the interactions between different sources of error in hybrid numerical-ANN solvers are lacking.

As an alternative to using \glspl{ann} to learn model errors, a smaller number of studies have pursued equation-discovery techniques~\cite{Lang_Tellus_2016, Mojgani_Chaos_2022, Pachev_SISC_2022}. In these approaches, one learns a closed-form, parsimonious representation of the analysis increment (an estimate of the model error) in terms of the state vector. For example in \citeA{Mojgani_Chaos_2022}, we introduced \gls{medida}. Such closed-form, parsimonious representations can be highly interpretable and point to the root cause of the model errors. They are also often generalizable, as they explain {the} underlying physical mechanisms and, in principle, can be connected to the physics of the changing system. This approach is also data efficient, usually requiring a small training set \cite{Lang_Tellus_2016, Zanna_GRL_2020, Mojgani_Chaos_2022, Ross_JAMES_2023, Jakhar_arxiv_2023}. While such model error representations might also lead to instabilities when used to correct the imperfect models, in principle, the existing rigorous tools and concepts for analyzing the stability of numerical methods can be applied to them. Finally, there are potential advantages of such closed-form equations over ANNs in terms of implementation and coupling to traditional numerical solvers.

In \citeA{Mojgani_Chaos_2022}, the capabilities of \method\ for discovering nonlinear structural errors were showcased on a highly chaotic, multi-scale canonical test case, i.e., the \gls{ks} equation. In that work, equation-discovery was performed using \gls{rvm}~\cite{Tipping_JMLR_2001}, which is a Bayesian linear regression with an interpretable library of linear and nonlinear bases built on the knowledge of domain experts and common physical terms (note that any equation-discovery method can be used in MEDIDA). Furthermore, it was shown that even if observations of the true system are noisy, \gls{da} methods such as \gls{enkf} can be used as smoothers with \method\ to successfully discover model errors.  

In this \doc, we first show the capabilities of \method\ using a much more complex and climate-relevant test case, the two-layer \gls{qg} turbulence. Second, we scale \method\ to work with spatially sparse observations, a challenging task that is essential for using model error discovery methods with imprecise observations such as those from in-situ and remote-sensing measurements. A critical component needed for dealing with sparse observation is an ``interpolator''. The quality of the interpolation scheme in equation discovery plays a significant role, as inaccuracies can degrade the library of bases that have high-order derivatives and/or strong nonlinearities. 

In the context of equation discovery, algebraic interpolators, such as polynomials~\cite{Rudy_SA_2017} and splines~\cite{Sun_Neurips_2022}, have shown success for canonical \gls{ode} or non-chaotic~\gls{pde} test cases. More recently, \glspl{ann} have emerged as popular and powerful interpolators to reconstruct the state and discover the governing equations from spatial sparsity~\cite{Chen_NatC_2021} or partially observed states~\cite{Lu_NatCP_2022}. Interpolation using \glspl{ann} provides a differentiable learning process that provides a flexible framework to formulate different optimization paradigms~\cite{Chen_NatC_2021}, an encoder to construct partial observations~\cite{Lu_NatCP_2022}, or to simultaneously estimate noise and the governing equation~\cite{Kaheman_MLST_2022}. However, the success of both algebraic and \gls{ann}-based interpolation schemes for highly chaotic and multi-scale test cases remains underexplored. In the case of \glspl{ann}, an implicit assumption is that the state and its higher order derivatives, e.g., dispersion or dissipation terms in \glspl{pde}, can be accurately represented by \glspl{ann}. This assumption is often justified by simply referring to \glspl{ann} as universal approximators~\cite{Hornik_NN_1989}.

In this \doc, we show that for highly multi-scale systems such as turbulent flows, \glspl{ann} do not learn the small-scale features (high-wavenumber features), a well-known phenomenon in the ML-literature called ``spectral-bias''~\cite{Rahaman_PMLR_2019,JohnXu_arxiv_2022}. This bias, which does not show up in simple metrics such as pattern correlation and \gls{rmse}, can cause significant errors in the calculations of high-order derivatives, hindering the successful equation discovery (of model error, and more broadly, other functions). We further show that this spectral bias can be mitigated by employing ``\glspl{rff}''\cite{Tancik_neurips_2020}, enabling the successful discovery of model errors from sparse observations.



Our contributions in this \doc\ are
\begin{enumerate}
    \item We scale \gls{medida} up to a much more complex and climate-relevant test case, the \gls{qg} turbulence,
    \item We extend the use of \gls{medida} to spatially sparse observations,
    \item We identify spectral bias as a major shortcoming of using common \glspl{ann} as interpolator for equation discovery in widely multi-scale systems such as QG turbulence,
    \item We show how this spectral bias can be alleviated using \glspl{rff}, enabling successful model error discovery from sparse observation where both algebraic interpolation and the naive use of \glspl{ann} fail, and 
    \item We present an example, based on missing hyperdiffusion, showing that short-horizon \gls{da} increments may not be able to capture some structural errors, although such errors may have significant long-term effects.
\end{enumerate}
Note that (3)-(4) have implications well beyond the topic of this study (model error, equation discovery) as spectral bias can significantly degrade the performance of \glspl{ann} when applied to multi-scale processes in the Earth system, leading to major challenges~(e.g., long term stability of the learned model~\cite{chattopadhyay2023long}).

This \doc\ is organized as follows.
\Cref{sec:method} summarizes the problem setup, our model error discovery framework, \method, and the new extensions to address spatial sparsity of observations. 
The details of our numerical experiments, i.e., the \gls{qg} flow (as the perfect system), imperfect test cases, and the library of bases are described in \cref{sec:experiments}.
The results on model error discovery from both fully observed and sparse observations are presented in \cref{sec:Results}. 
\Cref{sec:Conclusions} provides a summary and discussion.

\section{Model Error Discovery and Extensions to Sparse Observations}
\label{sec:method}

In this section, the elements of \method\ are briefly described. Additionally, \cref{sec:sparse} presents our proposed approach to estimate the full state from spatially sparse observations.

\subsection{Model error and state increment}
\label{sec:modelerror}
Here, the notion of {\it system} (the truth), {\it model} (the imperfect representation of the system), and {\it model errors} (differences between the system and model) are formalized. 
The goal is to discover model errors from sporadic and possibly sparse/noisy observations of the true system. Note that hereafter, ``system'' and ``observation'' will be used to refer to either the natural system (e.g., the real atmosphere) and the associated observations, or a high-fidelity simulation of this system (e.g., global 1-km runs) and the associated numerical results. 

Suppose that the spatio-temporal nonlinear evolution of the system is governed by a \gls{pde},
\begin{eqnarray}\label{eq:sys_truth}
	\partial_t \modelstateparc =  \funtruth{\modelstateparc},
\end{eqnarray}
where $\modelstateparc$ is the {state vector} in space and time.
We assume that the state can be observed in short temporal intervals, while the governing PDE operator, $\funtruth{\modelstateparc}$, is unknown to us. 
Moreover, we assume to have a model of the system described as
\begin{eqnarray}\label{eq:sys_model}
	\partial_t \modelstateparc =  \funmodel{\modelstateparc}.
\end{eqnarray}
We further assume to have a working numerical solver for the model, such that given an initial state, $\statemodel{}\left(x,\timepre_i\right)$, the state at the next time step, $\statemodel{}\left(x,\timenow\right)$, is predicted.

\begin{figure}[!t]
	\centering
	\def\myscale{0.42}
	\includegraphics[trim={0 120pt 0 0},clip,scale=\myscale]{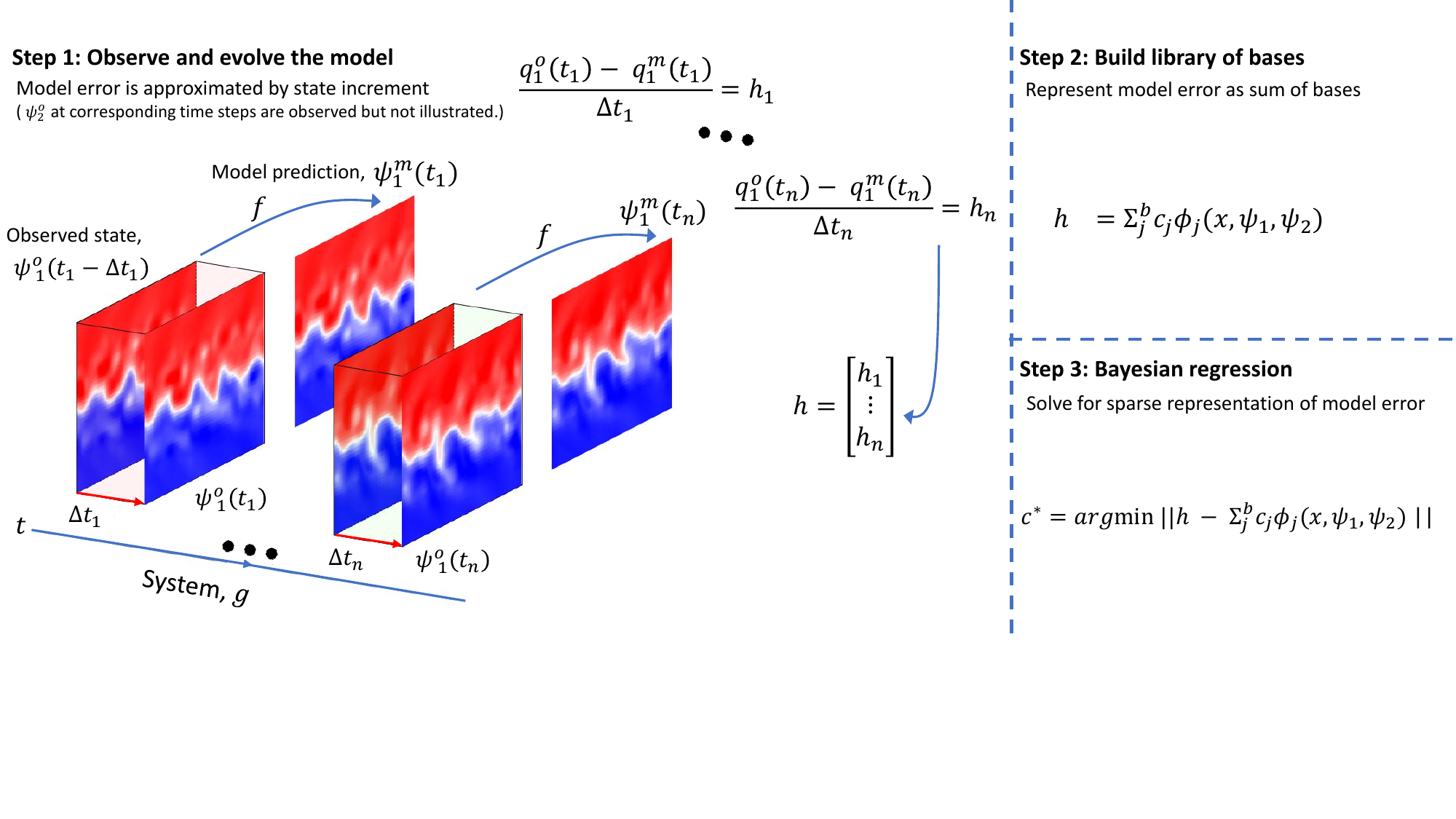}
	\caption{
		Schematic of the main steps of \method\ applied to the two-layer \gls{qg} test case. The states, $\psi_1$ and $\psi_2$, are observed at (potentially non-uniform) consecutive time steps in each later and the model error (in potential vorticity, $q$) in each layer is approximated from the analysis increment (Step 1). For clarity, only layer~1 is shown. A library of $b$ basis functions is constructed using the observations (Step 2) and the Bayesian regression problem is formulated and solved to discover the closed-form model error (Step 3). For better illustration, the schematic shows using only two pairs of snapshots; $h_i$ from several snapshots are stacked to form $h$.
		Calculations of $h$ and steps~2-3 are repeated for the discovery of the model error in layer 2.}
	\label{fig:MEDIDA}
\end{figure}

Without the loss of generality, the model error $\funerror{\modelstateparc}$ \revt{is assumed to be expressive} in an additive form~\cite{Levine_arxiv_2021, Mojgani_Chaos_2022}, i.e.,
\begin{eqnarray}\label{eq:operator_correction}
	\funerror{\modelstateparc}  = \funtruth{\modelstateparc} - \funmodel{\modelstateparc}.
\end{eqnarray}
With these definitions in mind, the goal of \gls{medida} is to approximate {structural and parametric} model error at time $t_i$, $h_i$, by leveraging one or more observation pairs, $w^o(x,t_i-\Delta t_i)$ and $w^o(x,t_i)$.
As described in \citeA{Mojgani_Chaos_2022} and shown here in Fig.~\ref{fig:MEDIDA} in the context of the QG test case, in \gls{medida}, the numerical solver of the model is initialized from the observed state, $w^o(x,t_i-\Delta t_i)$, to predict the state at $\statemodel{}\left(x,\timenow\right)$. The model error can then be estimated via the analysis increment as
\begin{eqnarray}\label{eq:diffu_clean}
	\begin{aligned}
		h_i(x)
		\approx
		(\stateobs{}\left(x,\timenow\right) - \statemodel{}\left(x,\timenow\right))/\timestep_i.
	\end{aligned}
\end{eqnarray}
With precise observations, $h_i$ includes structural/parametric model errors as well as errors from the numerical solver. 
\revt{Here, the structural/parametric model errors originate from the poor understanding of the true physical system (epistemic uncertainty), and numerical errors are results of discretization in time and/or space.}
As our goal is to isolate and quantify the former, the latter has to be minimized, e.g., by choosing small $\timestep_i$ and high spatial numerical resolutions. Treating observations as ``precise'' is reasonable if they are from high-fidelity simulations, but not if they are from measurements (in situ, remote sensing, etc.). In this case, the observations can be imprecise due to noise and/or spatial sparsity, or unavailability of some of the state variables, the so-called hidden or latent variables (i.e., partial state vector). In \citeA{Mojgani_Chaos_2022}, we showed that noisy observations can be used in \gls{medida} by employing \gls{da} techniques such as \gls{enkf}. In the current paper, we aim to tackle the challenge of spatial sparsity, as discussed in \cref{sec:sparse}. Note that in the current paper, we will not address the challenge of partial state vectors; see \citeA{Majda_Entropy_2018} and \citeA{Changhong_arxiv_2022} for recent advances in that area.     

\subsection{Interpretable model error discovery}
\label{sec:MED_lib}
The procedure in \cref{eq:diffu_clean} can be repeated given the availability of observation pairs to collect $\numsamples$ samples of $h_i$. Note that the time intervals, $\Delta t_i$, between the observation pairs do not have to be the same (but need to be short enough to reduce the numerical discretization errors). We assume that the \modelerror\ $h_i(x)$ can effectively be represented by a pre-defined library of basis functions $\phi_j$: 
\begin{eqnarray}
	h_i=  
	\Sigma_{j}^{\numlibcol} \coeffsymbole_j \basis_j \left(\grid, w(x,t_i)\right),
	\label{eq:model_error}
\end{eqnarray}
where $\coeffsymbole_j$ is the coefficient corresponding to the $j^\textit{th}$ term. The bases, $\phi_j$, can be {independent of the state}, e.g., to represent external forcings. However, they are often functions of the state variables, e.g., the spatial derivatives, polynomials, and their linear or nonlinear combinations~\cite{Rudy_SA_2017, Zanna_GRL_2020, Mojgani_Chaos_2022, Jakhar_arxiv_2023}. Such terms are usually chosen based on the understanding of the underlying physics~\cite{Lang_Tellus_2016, Zanna_GRL_2020, Jakhar_arxiv_2023}. That said, there are algorithms that provide more expressive and evolving libraries, e.g., through radial bases in a Gaussian process or symbolic regression, though the full physical interpretation of the discovered models via such algorithms can be complicated~\revt{\cite{Luo_Plos_2019, Levine_arxiv_2021, Chen_PhysicaD_2022, Ross_JAMES_2023}}.

Note that in this \doc, all calculations, including the derivatives for the library, are performed using the same Fourier spectral method used to solve the governing equations of the system and the model (see \cref{sec:neuralfitobs}).

\label{sec:MED_opt}
By assuming the candidate structural forms of the model error as in \eqref{eq:model_error}, we have reduced the problem of discovery of structural error into a parameter estimation problem, where the coefficients have to be estimated such that the difference between the prediction and observed state is minimized, i.e., a regression problem as in
\begin{linenomath*}
	\begin{equation}\label{eq:model_cost}
		\begin{aligned}
			\coeffdiscovered =~
			& \underset{ \coeff }{arg\minimize}
			\NormT{
				h - 
				\Sigma_{j}^{\numlibcol} \coeffsymbole_j \basis_j (x,w) 
			}, \\
		\end{aligned}
	\end{equation}
\end{linenomath*}
where $\coeffdiscovered = \left[\coeffsymbole^*_1, \coeffsymbole^*_2, \cdots, \coeffsymbole^*_\numlibcol \right]$ is the minimizer vector. Here, the $\numsamples$ samples of vector $h_i(x)$ are stacked to form the vector $h$.

Substituting the discovered model error in \eqref{eq:operator_correction} leads to the corrected model, i.e., 
\begin{eqnarray}\label{eq:sys_corrected}
	\partial_t \modelstateparc =  \funmodel{\modelstateparc}
	+
	\Sigma_{j}^{\numlibcol} \coeffsymbole^*_j \basis_j (x,w) .
\end{eqnarray}
To enhance interpretability, it is often favorable to limit the discovered model to its simplest form that can explain the highest variability, i.e., a parsimonious or sparse model~\cite{Rudy_SA_2017}. Enforcing sparsity is the cornerstone of equation discovery and many interpretable modeling efforts in statistical learning~\cite{hastie2015statistical}. In this \doc, parsimony is achieved using \gls{rvm}, a Bayesian regression algorithm that has been successfully applied to various \gls{ode}/\gls{pde} discovery problems \cite{Zhang_RSP_2018,Zanna_GRL_2020,Mojgani_Chaos_2022,Jakhar_arxiv_2023}.
A threshold on the variance of fitted coefficients is systematically chosen to prune the less relevant terms (with the highest uncertainty in the coefficient).
\revt{In this procedure, a range for the model hyper-parameter is swept leading to a Pareto front, and the elbow of the L-curve is chosen as a compromise between accuracy and sparsity of the model.}
See~\citeA{Zhang_RSP_2018} and \revt{\citeA{Jakhar_arxiv_2023}} for more details and discussions of \gls{pde} discovery using \gls{rvm}.


The accuracy of corrected models is quantified using the normalized distance between the coefficients of the model and the true system~\cite{Rudy_SA_2017, Reinbold_PRE_2020, Mojgani_Chaos_2022}:
\begin{linenomath*}
	\begin{subequations}\label{eq:error_coeff}
		\begin{align}
			\errorcoeffmodel &= 
			\frac{ \NormT{ \coefftrue - \coeffmodel }}{ \NormT{ \coefftrue } } \times 100,
			\\
			\errorcoeffdiscovered &= \frac{ \NormT{ \coefftrue - \coeffdiscovered }}{ \NormT{ \coefftrue}} \times 100,
		\end{align}
	\end{subequations}
\end{linenomath*}
where $\coefftrue$ is the vector of the coefficients in the true system, and $\coeffmodel$ and $\coeffdiscovered$ are the vectors of coefficients in the model and corrected model, respectively.
\revt{We acknowledge that this measure is not practical when the true system is unknown, however, this is commonly used measure to validate a methodology, see., e.g.,~\citeA{Rudy_SA_2017, Reinbold_PRE_2020}. In practice, the predictive capability of the discovered model can be evaluated after the discovery step, an \aposteriori\ test.}
The vector of coefficients span over the library of bases defined in \cref{sec:MED_lib}, i.e., they are of size $\numlibcol$.
A successful discovery leads to $\errorcoeffdiscovered \ll \errorcoeffmodel$.
Note that although this measure is often used in the equation-discovery literature, it can also be misleading when a parameter of a dynamically important term is incorrect but is much smaller than some of the other coefficients, or when such a term is entirely missing (structural error). In such cases, long-term simulations of the model can be statistically very different from those of the system even when $\errorcoeffmodel$ is small, thus $\errorcoeffdiscovered \ll \errorcoeffmodel$ might fail as a proper critetion. Therefore, examining the statistics of the corrected model would be needed in such cases to provide a better metric. In our study, given that we precisely know the governing equations of the system, it suffices to examine the individual coefficients, as presented in the results section.

\subsection{Interpolation and discovery from spatially sparse observations}
\label{sec:sparse}

The previous steps assume the availability of observations of the full state. 
However, observations, e.g., in-situ or from remote sensing, are often sparse in the spatial domain. Therefore, to compute $h_i$ and $\phi_j$ in \gls{medida}, we need an additional ``interpolation'' step to estimate the full observed state. Then, the estimated state as the output of this step will readily, without requiring any further modifications to the framework, replace the fully observed states in Eqs.~\eqref{eq:model_error}-\eqref{eq:sys_corrected}.

As shown later, the quality of this interpolation plays a major role in the success of equation discovery, in particular, because of the presence of high-order spatial derivatives in the library of $\phi_j$. Combining prior information with sparse observations to estimate/re-construct the full state is known as the state estimation step and is a key component of \gls{da} algorithms such as Kalman filters/smoothers~\revt{\cite{evensen2022data}}. However, interpolation is known to be particularly challenging for multi-scale systems such as turbulent flows~\cite{Willcox_CF_2006, Shengyu_IEEE_2021, Beauchamp_arxiv_2022, DiLeoni_arxiv_2022, Kelshaw_arxiv_2022, Karnakov_arxiv_2023}. In fact, our attempts in using \gls{enkf} using algebraic interpolations showed that an impractically large ensemble size might be required to achieve the estimation accuracy needed for a successful equation discovery. We emphasize again that the requirement for high accuracy in the interpolation task comes from the need for constructing an accurate library of high-order derivatives and strong nonlinearities. Another approach to state estimation/reconstruction that has received substantial attention in recent years involves the use of \glspl{ann} as interpolators. This approach has shown success in a number of recent studies involving multi-scale toy models, including those motivated by Earth system-related applications~\cite{Chen_NatC_2021, Beauchamp_arxiv_2022}.    

\begin{figure}
	\centering
           \includegraphics[trim={0 0 0 0},clip]{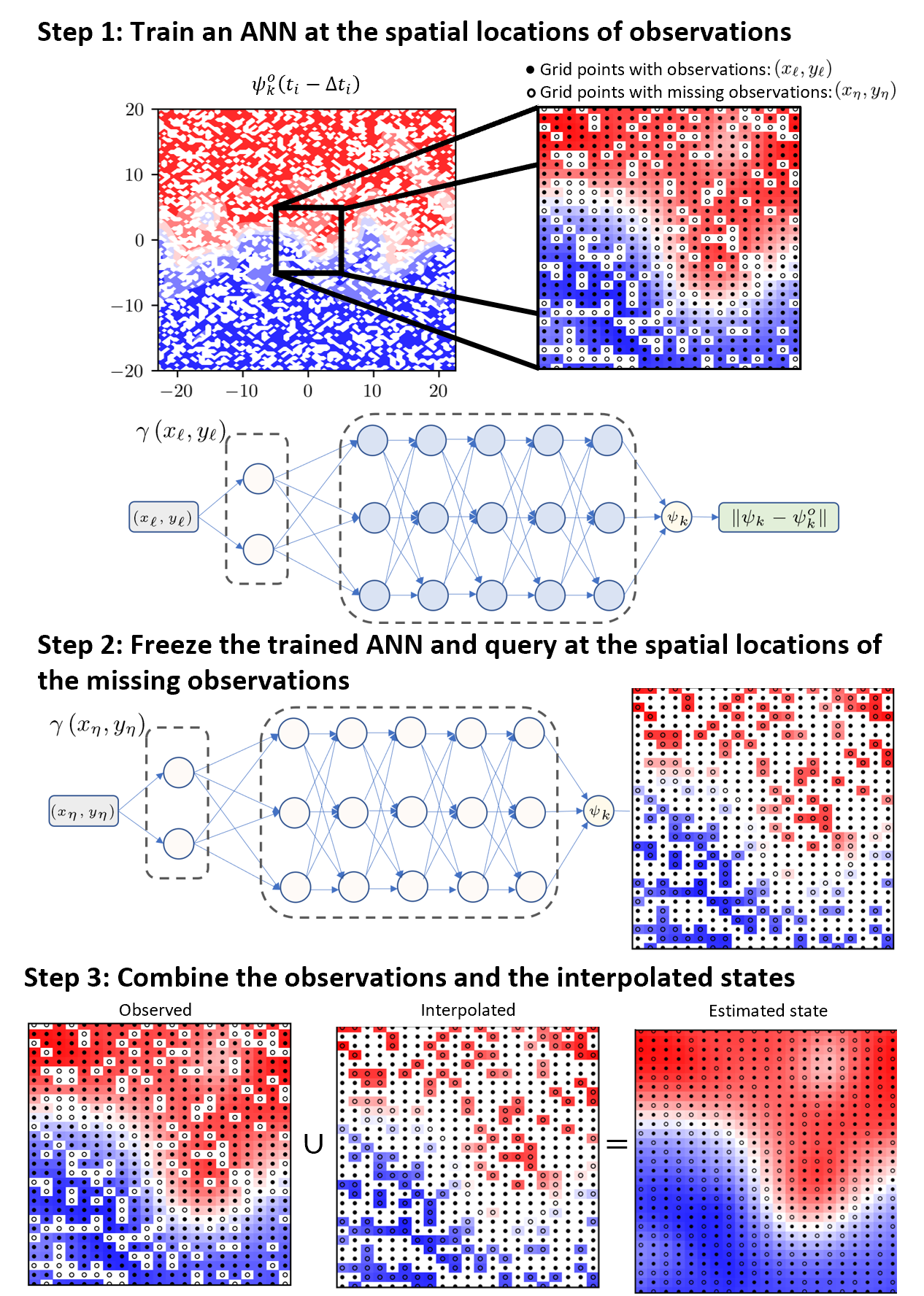}
	\caption{\revt{
	Schematic of the training process of the \gls{ann}-based interpolator on the observations, followed by the inference step at the location of the missing observations. An \gls{ann} is trained at the location of the known observation datapoints. 
	After the training phase is complete, the trained model is queried at the location of the missing observations. The union of the known datapoints and the interpolated states provides us with an estimation of the state necessary for the other steps of \method. 
	}}
	\label{fig:schematic_interp}
\end{figure}

In this \doc, we leverage these advances and use \gls{ann}-based interpolators. Details of these ANN-intepolators are given in \cref{sec:interp}. Briefly, for a given snapshot at time $t_i$ on a grid of size $L$, suppose $\ell=1, 2, \dots, L-S$ is the index of grid points at which the value of a function $A(x_\ell,y_\ell)$ is known (i.e., observation datapoints), and $\eta=1, 2, \dots, S$ is the index of grid points at which the value of this functions is unknown (the level of sparsity is $S/L$). We train an ANN (a 5-layer perceptron) with inputs of $(x_\ell,y_\ell)$ and outputs of $A(x_\ell,y_\ell)$ and a mean-square-error loss function. Once trained, the ANN estimates the missing values $A(x_\eta,y_\eta)$ for inputs of $(x_\eta,y_\eta)$. 
\revt{This procedure is illustrated in~\cref{fig:schematic_interp}.}
Note that unlike many other data-driven interpolators~\cite{Willcox_CF_2006,Fukami_NMI_2021,Beauchamp_arxiv_2022}, this approach does not need a history of past sparse observations, although using past data may become beneficial when the level of sparsity increases. As a baseline for our comparisons, we use an algebraic (cubic) interpolator trained on the same observations. 

To summarize, the trained interpolators (either algebraic or \gls{ann}-based) for each of the state parameters are queried at the location of missing observations to estimate a full state at $\timepre_i$.
This estimation is then evolved using the model \eqref{eq:sys_model} to predict  $\statemodel{}\left(\timenow\right)$.
Moreover, the full observed state at the following time step, $\stateobs{}\left(\timenow\right)$, is  similarly estimated from sparse  observations.
The approximated states are used to construct the library of bases in \eqref{eq:model_error}, and finally the sparse regression problem of model error \eqref{eq:model_cost} is solved.

Note that our use of \glspl{ann} is restricted to interpolation, and unlike many studies mentioned in \cref{sec:introduction}, the \glspl{ann} are not used to learn the model error. 




\section{The Quasi-Geostrophic System}
\label{sec:experiments}
\def\friclin{ \nabla^{2} \psi_{2} / \tau_{f} }
\def\fricquadratic{\partial_x\left(\left|{\nabla} \psi_2\right| \partial_x \psi_2\right)+\partial_y\left(\left|{\nabla} \psi_2\right| \partial_y \psi_2\right) }
\newcommand{\fric}{\ensuremath{ f }}
\newcommand{\QGoneRHSt}{\ensuremath{ 
		- \alpha J\left(\psi_{1}, q_{1}\right) + 
		\frac{1}{ \tau_{d_1}}\left(\psi_{1}-\psi_{2} \right)
		-\frac{1}{ \tau_{d_2}}\left( \psi_{R}\right)   
}}
\newcommand{\QGtwoRHSt}{\ensuremath{ 
		- \alpha J\left(\psi_{2}, q_{2}\right) 
		- \frac{1}{ \tau_{d_1} }\left(\psi_{1}-\psi_{2} \right) 
		+ \frac{1}{ \tau_{d_2} }\left(\psi_{R}\right) 
		- \fric
}}
\newcommand{\QGkRHSt}{\ensuremath{ 
		- \alpha J\left(\psi_{k}, q_{k}\right) 
		- \frac{ \left(-1\right)^k}{ \tau_{d_1} }\left(\psi_{1}-\psi_{2} \right) 
		+ \frac{ \left(-1\right)^k }{ \tau_{d_2} }\left(\psi_{R}\right) 
		- \delta_{k2} \fric
}}

\newcommand{\QGonec}{\ensuremath{
		c_7 J\left(\psi_{1}, q_{1}\right) 
		+ c_4 \psi_{1} + c_5 \psi_{2}
		+ c_6 \psi_{R} 
		+ c_0 \nabla^{2} \psi_{1} + c_1 \nabla^{2} \psi_{2}
}}
\newcommand{\QGtwoc}{\ensuremath{
		c_8 J\left(\psi_{2}, q_{2}\right)
		+ c_4 \psi_{1} + c_5 \psi_{2} 
		+ c_6 \psi_{R} 
		+ c_0 \nabla^{2} \psi_{1} + c_1 \nabla^{2} \psi_{2}
}}

\newcommand{\QGtk}{\ensuremath{
		c_1 \nabla^{2} \psi_{1} 
		+ c_2 \nabla^{2} \psi_{2}
		+ c_3 \nabla \psi_{1} 
		+ c_4 \nabla \psi_{2}
		+ c_5 \psi_{1}
		+ c_6 \psi_{2}
		+ c_7 \psi_{R}
		+ c_8 J\left(\psi_{1}, q_{1}\right) 
		+ c_9 J\left(\psi_{2}, q_{2}\right)
}}

\newcommand{\QGtkshort}{\ensuremath{
		c_0 \nabla^{2} \psi_{1} 
		+ c_1 \nabla^{2} \psi_{2}
		+ c_2 \psi_{1}
		+ c_3 \psi_{2}
		+ c_4 \psi_{R}
		+ c_5 J\left(\psi_{1}, q_{1}\right) 
		+ c_6 J\left(\psi_{2}, q_{2}\right)
}}
\subsection{The true system}
\label{sec:truesystem}

As the true system, consider the non-dimensionalized two--layer \gls{qg} model \cite{Phillips_T_1954}, a simple baroclinic climate model: 
\begin{linenomath*}
	\begin{subequations}\label{eq:qg}
		\begin{align}
			\frac{\partial q_{1}}{\partial t} &= \QGoneRHS, \\
			\frac{\partial q_{2}}{\partial t} &= \QGtwoRHS,
		\end{align}
	\end{subequations}
\end{linenomath*}
where subscripts $1$ and $2$ represent the upper and lower layers, respectively.
Here, $q$ is the potential vorticity, and $\streamfunction{}$ is the stream function.
Equilibrium profile, $\psi_R$, is defined to enforce a baroclinically unstable jet, and is set to $\partial \psi_R / \partial y = \sech^2 \left( y/\sigma\right)$, where $\sigma$ is the width of the jet.
$\tau_f$ is the Rayleigh friction time scale in the term that represents surface drag. $\tau_{{d}_1}$ and $\tau_{{d}_2}$ are the Newtonian relaxation time scales in the term presenting radiative cooling.
$\nu$ is the hyperdiffusion coefficient in the term that represents unresolved physics.
The Jacobian term is 
\begin{linenomath*}
	\begin{subequations}
		\begin{align}\label{eq:jacobian}
			J(\psi_k, q_k) =\alpha \left( \frac{\partial q_k}{\partial x} \frac{\partial \psi_k}{\partial y} - \frac{\partial q_k}{\partial y} \frac{\partial \psi_k}{\partial x} \right),
		\end{align}
	\end{subequations}
\end{linenomath*}
for both layers, $k\in\left\{1,2\right\}$, and $\alpha$ is a constant we introduce to impose a model error in the nonlinear terms (for the true system $\alpha=1$). Potential vorticity in each layer is
\begin{linenomath*}
	\begin{subequations}\label{eq:qg_q}
		\begin{align}
			q_1 &= \nabla^{2} \streamfunction{1} - \streamfunction{1} + \streamfunction{2} + \beta y,  \\
			q_2 &= \nabla^{2} \streamfunction{2} + \streamfunction{1} - \streamfunction{2} + \beta y + R \left(x,y\right),
		\end{align}
	\end{subequations}
\end{linenomath*}
where $\beta$ is the meridional gradient (in $y$-direction) of the Coriolis parameter, and $R\left(x,y\right)$ is the orographic forcing only acting on the lower layer. 
The orography is represented by the summation of Gaussian features, each centered at $\left[{x_i},{y_i}\right]$ with the height of $r_i$, and variance of $\sigma_i^2$, i.e.,
\begin{linenomath*}
	\begin{equation}\label{eq:forcing}
		\begin{array}{c}
			R\left(x,y\right) = \sum_{i=1}^{N_r}\topography{r_i}{x_i}{y_i}{\sigma_i^2}.
		\end{array}
	\end{equation}
\end{linenomath*}
Equations~\eqref{eq:qg} are integrated using the pseudo-spectral solver of~\citeA{Lutsko_JAS_2015}.
All parameters are the same as those in~\citeA{Lutsko_JAS_2015}, and are summarized in \cref{tab:qg_cases}; note that $t=5= 200 \Delta t$ is $1$ Earth day. See \citeA{zurita2014impact}, \citeA{Lutsko_JAS_2015} and \citeA{Nabizadeh_GRL_2019} for detailed discussions and analyses of the system.
The non-dimensionalized zonal and meridional domain widths are $L_x=46$ and $L_y=68$\revt. $96\times192$ Fourier modes are used in the $x$ and $y$ directions, respectively.
The width of the jet is set to $\sigma=3.5$.
The boundary effects on the northern and southern ends are damped by a sponge layer. 
The orographic forcing, when present, is described by \eqref{eq:forcing} where  $\left(x_i,y_i\right)\in \left\{ \left(0,0\right), \left(-10,-5\right), \left(0,20\right), \left(0,-20\right)\right\}$, $r_i=20$, and $\sigma_i=5$ for $i\in\left\{1,2,3,4\right\}$ (depicted in \cref{fig:R_truth}).



\subsection{The imperfect models}
\label{sec:imperfectmodel}

For our numerical experiments, we have defined 12 (imperfect) models introduced in \cref{tab:qg_cases}.
These cases consist of a combination of parametric and structural uncertainties due to different terms, representing different physical phenomena. 
Cases 1 and 2 include parametric uncertainty on the radiative cooling, and both drag and radiative cooling terms. 
In Case 3, the radiative cooling term is completely missing, i.e., a structural uncertainty. 
Case~4 consists of a parametric uncertainty modeling the nonlinear Jacobian term. 
Case~5 includes a combination of parametric and structural uncertainties in all the aforementioned terms.
Case~6 has only a large parametric error in the drag term.
Cases 7 to 10 include structural uncertainties in different terms.
Cases 1-3 and 6-8 represent model errors in linear terms while Cases 4 and 5 have errors in the nonlinear terms. 

In Case~9, $\beta$ is missing, representing a major structural model error in the dynamics of the flow. In Case~10, the hyperdiffusion term is missing, another major structural error that as discussed later, does not dominate the analysis increment, as its effects are not significant in short-term evolution.

In Case~11, the drag term is modeled incorrectly. Inspired by~\citeA{Gallet_pnas_2020}, the drag term is assumed to have a quadratic form, i.e., 
\begin{linenomath*}
	\begin{equation}\label{eq:drag_quad}
		\begin{array}{c}
			\fric = -0.07  \left[\fricquadratic\right]
		\end{array}
	\end{equation}
\end{linenomath*}
and replaces the linear drag in layer 2. In this case, a successful discovery has to remove the extra terms and replace them with correct terms from the library of bases.

Finally, in Case~12, the orographic forcing is misrepresented, i.e., $R\left(x,y\right) \neq R_m\left(x,y\right)$ as shown in \cref{fig:R_truth} versus \cref{fig:R_model}. The model error consists of misrepresented amplitudes of the forcing terms, i.e., $\left(x_i,y_i\right)\in \left\{ \left(0,0\right), \left(-10,-5\right), \left(0,20\right), \left(0,-20\right)\right\}$, $r'_i=\left\{20,10,15,5\right\}$, and $\sigma_i'=5$ for $i\in\left\{1,2,3,4\right\}$. 
A successful discovery should identify the differences between the magnitudes of forcing terms from the library.

The twelve cases described above provide a variety of benchmarks where parametric and structural uncertainty and misrepresentation of the system are incorporated into the model.

\subsection{The library of bases}
\label{sec:libraryqg}
The library, $\basis_i$ as defined in \cref{sec:MED_lib}, consists of the candidate terms of the state variables and their first and second derivatives in both zonal and meridional directions, i.e.,

\begin{linenomath*}
	\begin{equation}
		\begin{aligned}
			\{
			\psi_R,
			\psi_i, 
			\frac{\partial \psi_i}{\partial x}, 
			\frac{\partial \psi_i}{\partial y}, 
			\psi_i \frac{\partial \psi_i}{\partial x}, 
			\psi_i \frac{\partial \psi_i}{\partial y}, 
			J(\psi_i, q_i),
			\frac{\partial^2 \psi_i}{\partial x^2}, 
			\frac{\partial^2 \psi_i}{\partial y^2}, 
			\frac{\partial^4 \psi_i}{\partial x^4}, 
			\frac{\partial^4 \psi_i}{\partial y^4}, 
			\frac{\partial^8 \psi_i}{\partial x^8}, 
			\frac{\partial^8 \psi_i}{\partial y^8}
			\},
			\quad i\in\left\{1,2\right\}.
		\end{aligned}
		\label{eq:library_QG}
	\end{equation}
\end{linenomath*}
noting that
$\nabla^{2} \psi_{k} = \frac{\partial^2 \psi_k}{\partial x^2} + \frac{\partial^2 \psi_k}{\partial y^2}$.
Any discovered orographic terms,  $r_i \left(x_i, y_i\right)$, are expected to appear in the nonlinear form of
\begin{linenomath*}
	\begin{equation}
		\begin{aligned}
			\left\{
			\frac{\partial \psi_{2}}{\partial y} \frac{\partial R_i}{\partial x} ,
			\frac{\partial \psi_{2}}{\partial x} \frac{\partial R_i}{\partial y} 
			\right\}, i \in \left\{1,2,3,\cdots,25\right\}
		\end{aligned}
	\end{equation}
\end{linenomath*}
where the candidate bases, $R_i(x_i,y_i)=\topography{r_i}{x_i}{y_i}{\sigma_i^2}$ , are uniformly distributed on a $5\times5$ formation in the domain with variance of $\sigma_i=5$.
Moreover, the following nonlinear terms are included to represent the quadratic drag terms,
\begin{linenomath*}
	\begin{equation}
		\begin{aligned}
			\left\{
			\partial_x \left( \left| \nabla \psi_i \right| \partial_x \psi_i \right), 
			\partial_y \left( \left| \nabla \psi_i \right| \partial_y \psi_i \right)
			\right\},
			i,j\in\left\{1,2\right\},
		\end{aligned}
	\end{equation}
\end{linenomath*}
and to enrich the space of exploration, we also include
\begin{linenomath*}
	\begin{equation}
		\begin{aligned}
			\left\{
			\frac{\partial \psi_{i}}{\partial y} \frac{\partial \psi_{j}}{\partial x} 
			,
			\frac{\partial \psi_{i}}{\partial y} \frac{\partial \nabla^{2} \psi_{j}}{\partial x}
			,
			\frac{\partial \psi_{i}}{\partial x} \frac{\partial \nabla^{2} \psi_{j}}{\partial y} 
			\right\},
			i,j\in\left\{1,2\right\}.
		\end{aligned}
	\end{equation}
\end{linenomath*}
This choice of the library results in $85$ candidate terms, consisting of linear and nonlinear terms.

\section{Numerical Experiments}
\label{sec:Results}

In the following experiments, the first $500$ Earth days of the simulation of the true system are dismissed as the spin-up period. Then, $n=100$ pairs of observations $(\psi_k^o(x,\timepre), \psi_k^o(x,\timenow))$ for $k=\left\{1,2\right\}$ are collected daily from a $100$-day training period. The interpolation step (described in \cref{sec:sparse}) is applied if the observations are sparse. Next, each imperfect model (described in \cref{sec:imperfectmodel}) is evolved for one $\timestep$ from each $\psi_k^o(x,\timepre)$ and the analysis increment is calculated. 
The library of bases is constructed as in~\cref{sec:libraryqg}, and the minimization problem for discovery is solved as described in \cref{sec:MED_opt}.

In \cref{sec:fullobs}, \gls{medida} is evaluated for when full observations of the state are available (referred to as {\it direct}). In \cref{sec:neuralfitobs}, the same full observations are used to train \glspl{ann} to evaluate the capabilities of \glspl{ann} as interpolation schemes. As shown soon, we find spectral bias as a major shortcoming of such interpolators for equation discovery, which we will then address using \glspl{rff}. Then, in \cref{sec:missingobs}, \gls{medida} with ANN+RFF interpolators (and baseline algebraic interpolators) is evaluated for spatially sparse observations.

\subsection{Discovery from full observations (Direct)}
\label{sec:fullobs}
First, we evaluate \gls{medida} given the full observation of the state variables, $\psi_1$ and $\psi_2$. Here, the analysis increments $h_i$ and library terms are constructed  {\it directly} using measurements of the state variables. The performance of \method\ for the first $10$ test cases of parametric and structural errors (\cref{tab:qg_cases}) is quantified and summarized in \cref{tab:qg_result}. In all the first 9 cases, which involve linear and nonlinear structural errors, the proposed approach has reduced the model error significantly: $\varepsilon_m=\order{10}-\order{100}\%$ has been reduced to $\errorcoeffdiscovered=\order{0.1}-\order{0.01}\%$ (the one exception is Case 5, layer 2, where $\varepsilon_m=103.6\%$ is reduced to $\errorcoeffdiscovered=1.3\%$, which is still a substantial reduction). For selected representative cases, Fig.~\ref{fig:discovery_selected} compares the true, model's, and corrected model's coefficients for some of the library terms (note the logarithmic scale). We see that in all these cases, \method\ successfully corrects the model (compare the black, red, and blue bars); this includes a case where the entire Jacobian in layer~2 is missing (\cref{fig:discovery_selected}-(\subref{fig:c5l2})).


Case~10 is the only unsuccessful discovery:  In this case, the model error is missing scale-selective hyperdiffusion, which is needed to remove enstrophy from small scales and ensure the long-term stability of the solver. However, the effects of missing hyperdiffusion are small in short-term evolutions and therefore, do not dominate the analysis increment (numerical errors, for example, could be comparable or even larger in magnitudes). As a result, \method\, or any method that relies on model error discovery from analysis increment, will fail. This case serves as an important reminder of this shortcoming of such approaches, as there are important processes that have small effects on short-term evolution but significant effect on the long-term evolution of the climate system (one example is cloud microphysics~\cite{atlas2023tropical}). 


For Cases~11-12, the model error discovery is successful. For Case~11, where the structural error is quadratic surface drag (Eq.~\eqref{eq:drag_quad}) instead of a linear drag ($-0.07 \nabla^2 \psi_2$), \method\ successfully finds the following correction 
\begin{linenomath*}
	\begin{equation}
		h = + 0.07  \left[\fricquadratic\right] - 0.0704 \nabla^2 \psi_{2},    
	\end{equation}
\end{linenomath*}
which removes the incorrect quadratic term and adds the linear drag with only $0.6\%$ error in $\tau_{f}$ ($\errorcoeffdiscovered<0.01 \%$).

Finally, in Case~12, where the orography's profile $R(x,y)$ is incorrect (compare \Cref{fig:R_model} with \Cref{fig:R_truth}), \method\ successfully corrects the profile (\Cref{fig:R_corrected}), reducing the model error to $\errorcoeffdiscovered<0.01\%$.

As reported in \citeA{Mojgani_Chaos_2022} for a much simpler chaotic test case, here, we again find that \method\ can effectively and robustly discover complex and nonlinear structural errors if the state is fully observed and if the model error dominates the analysis increment. The next question is the performance of \method\ with sparse observations. As mentioned before, simply using a linear or cubic interpolation scheme did not lead to satisfactory discoveries in most cases (\ref{tab:qg_result}). The use of ANN-based interpolators resulted in even worse failures (not shown). To understand the source of these failures, next, we examine the performance of ANNs when simply used to ``represent'' the fully observed states.

\def\colorterm{black}

\newcommand\mytauf{\ensuremath{0.07}}
\newcommand\mytaud{\ensuremath{0.01}}

\newcommand\myjacmodel{\ensuremath{\fbox{\color{\colorterm} 0.5}}}

\newcommand\mytaufmodel{\ensuremath{\fbox{\color{\colorterm} 0.37}}}
\newcommand\mytaudmodel{\ensuremath{\fbox{\color{\colorterm} 0.11}}}
\newcommand\mytaufmodelt{\ensuremath{\underline{\color{\colorterm} 0.01}}}
\newcommand\mytaudmodelt{\ensuremath{\underline{\color{\colorterm} 0.02}}}

\newcommand\mybetamodel{{\color{\colorterm}0.006}}
\newcommand\mybeta{0.196}

\begin{table}[!tb]
	\caption{
		Twelve cases of imperfect models, which suffer from a range of parametric and/or structural errors due to different mis-represented physics.
		The coefficients of the system are in the top row ($\varepsilon_m=0$).
		Boxes are around values that represent parametric model errors (P). 
		The \mycros\ represents structural errors (S), i.e., missing terms.
		The error of the incorrect models, $\errorcoeffmodel (\%)$, is calculated separately for each layer.
	}
	\label{tab:qg_cases}
	\centering
	\resizebox{\textwidth}{!}{%
		\begin{tabular}{ccccccccc|cc}
			\hline
			\multirow{2}{*}{Case} & \multirow{2}{*}{Model error} & \multirow{2}{*}{Physics of model error} & \multicolumn{6}{c|}{System/model parameters} & \multicolumn{2}{c}{$\errorcoeffmodel (\%)$} \\ \cline{4-11} 
			&  &  & $\alpha$ & ${1}/{ \tau_{{d}_1}}$ & ${1}/{ \tau_{{d}_2}}$ & ${1}/{ \tau_{f}}$ & $\beta$ & $\nu$ & \multicolumn{1}{c}{Layer 1} & \multicolumn{1}{c}{Layer 2} \\ \hline
			& & None & 1 & 0.01 & 0.01 & 0.07 & 0.196 & $10^{-6}$ & 0 & 0 \\ \hline
			1 & P. & {Radiative cooling} & 1 & \mytaudmodel & \mytaudmodel & 0.07 & 0.196 & $10^{-6}$ & 16.99 & 16.95 \\
			2 & P. & {Radiative cooling} and drag & 1 & \mytaudmodel & \mytaudmodel & \mytaufmodel & 0.196 & $10^{-6}$ & 16.99 & 33.91 \\
			3 & S. & {Radiative cooling} & 1 & \mycros & \mycros & 0.07 & 0.196 & $10^{-6}$ & 1.7 & 1.7 \\
			4 & P. & Nonlinear term & \myjacmodel & 0.01 & 0.01 & 0.07 & 0.196 & $10^{-6}$ & 49.06 & 48.94 \\
			5 & P.+S. & \makecell{Nonlinear term, \\ {radiative cooling}, drag} & \mycros & \mytaudmodel & \mytaudmodel & \mytaufmodel & 0.196 & $10^{-6}$ & 99.58 & 103.59 \\
			6 & P. & {Drag} & 1 & 0.01 & 0.01 & \mytaufmodel & 0.196 & $10^{-6}$ & 0 & 29.37 \\
			7 & S. & {Drag} & 1 & 0.01 & 0.01 & \mycros & 0.196 & $10^{-6}$ & 0 & 6.85 \\
			8 & S. & {Radiative cooling} & 1 & 0.01 & \mycros & 0.07 & 0.196 & $10^{-6}$ & 0.98 & 0.98 \\
			9 & S. & Coriolis gradient & 1 & 0.01 & 0.01 & 0.07 & \mycros & $10^{-6}$ & 19.23 & 19.9 \\
			10 & S. & Hyperdiffusion & 1 & 0.01 & 0.01 & 0.07 & 0.196 & \mycros & $<0.01$ & $<0.01$ \\
			11 & S. & {Model of drag (see \cref{eq:drag_quad})} & 1 & 0.01 & 0.01 & 0.07 & 0.196 & $10^{-6}$ & 0 & $11.87$ \\
			12 & S. & {Orography (see  \crefrange{fig:R_truth}{fig:R_model})} & 1 & 0.01 & 0.01 & 0.07 & 0.196 & $10^{-6}$ & 0 & $39.52$ \\
			
			\bottomrule
		\end{tabular}
	}
\end{table}

\begin{table}[!tb]
	\caption{
		For the first 10 cases, the errors of the corrected models $\errorcoeffdiscovered (\%)$ versus the errors of the original imperfect model $\varepsilon_m (\%)$. The discovery is performed separately for each layer, and the errors are reported separately.
	}
	\def\fno{$^\dagger$}
	\def\fnt{$^\ddagger$}
	\def\fnth{$^\text{o}$}
	\def\myfail{ \multirow{17}{*}{\rotatebox[origin=c]{90}{$>100$ (Discovery failed)\fnth}}  }
	\def\myfailh{ \multicolumn{5}{c}{\multirow{2}{*}{$>100$ (Discovery failed)\fnt}}   }
	\newcommand\mycaset[1]{\multirow{2}{*}{#1}}
	\label{tab:qg_result}
	\centering
	\resizebox{\textwidth}{!}{%
		\begin{tabular}{cc|c|ccccc}
			\hline
			\multirow{3}{*}{Case}  & 
			\multirow{3}{*}{Layer} & 
			\multirow{3}{*}{$\varepsilon_m (\%)$} & 
			\multicolumn{5}{c}{\hspace{-40pt}$\errorcoeffdiscovered  (\%)$} \\
			&                        &                                & 
			\multicolumn{3}{c|}{Full observation (no sparsity)}                     &
			\multicolumn{2}{c}{Sparse observation ($5\%$ sparsity)} \\\cline{4-8} 
			&                        &                                & Direct    & ANN                    & \multicolumn{1}{c|}{ANN+RFF} & Cubic      & ANN+RFF       \\ \hline
			\mycaset{1}	&1 & $16.99$ & $0.04$ & \myfail & \multicolumn{1}{c|}{$0.31$}	&$10.92$ & $0.07$\\
			&2 & $16.95$ & $0.04$ &  		& \multicolumn{1}{c|}{$0.02$}	&$5.53$	 & $0.08$\\ \cline{1-4} \cline{6-8} 
			\mycaset{2} &1 & $16.99$ & $0.04$ &  		& \multicolumn{1}{c|}{$0.03$}	&$11.00$ & $0.61$\\
			&2 & $33.91$ & $0.19$ &  		& \multicolumn{1}{c|}{$0.26$}	&$6.26$	 & $1.04$\\ \cline{1-4} \cline{6-8} 
			\mycaset{3} &1 & $1.70$  & $<0.01$&  		& \multicolumn{1}{c|}{$2.20$} 	&$8.99$	 & $1.69$\\
			&2 & $1.70$	 & $<0.01$&  		& \multicolumn{1}{c|}{$0.21$}	&$6.39$	 & $1.76$\\ \cline{1-4} \cline{6-8} 
			\mycaset{4} &1 & $49.06$ & $0.13$ & 		& \multicolumn{1}{c|}{$0.64$}   &$6.13$	 & $0.71$\\
			&2 & $48.94$ & $<0.01$& 		& \multicolumn{1}{c|}{$0.13$} 	&$4.41$	 & $0.54$\\ \cline{1-4} \cline{6-8} 
			\mycaset{5} &1 & $99.58$ & $0.34$ &  		& \multicolumn{1}{c|}{$1.25$}	&$17.98$ & $0.96$\\
			&2 & $103.59$& $1.31$ &  		& \multicolumn{1}{c|}{$1.45$}	&$4.69$	 & $3.93$\\ \cline{1-4} \cline{6-8} 
			\mycaset{6} &1\fno & 0  & 0	  &  		    & \multicolumn{1}{c|}{$0.06$} 	&$0.15$  & $0.10$ 	\\
			&2 & $29.37$ & $0.14$ &  		& \multicolumn{1}{c|}{$0.08$}	&$5.41$	 & $2.09$\\ \cline{1-4} \cline{6-8} 
			\mycaset{7} &1\fno & 0  & 0	  &  		    & \multicolumn{1}{c|}{$0.05$} 	&$0.12$  & $0.09$ 	\\
			&2 & $6.85$  & $<0.01$&  		& \multicolumn{1}{c|}{$0.04$}	&$6.96$	 & $1.62$\\ \cline{1-4} \cline{6-8} 
			\mycaset{8} &1 & $0.98$  & $<0.01$&  		& \multicolumn{1}{c|}{$2.10$} 	&$2.18$	 & $0.04$\\
			&2 & $0.98$  & $<0.01$&  		& \multicolumn{1}{c|}{$<0.01$} 	&$6.33$	 & $0.23$\\ \cline{1-4} \cline{6-8} 
			\mycaset{9} &1 & $19.20$ & $0.02$ &  		& \multicolumn{1}{c|}{$0.02$} 	&$0.02$	 & $0.02$\\
			&2 & $19.19$ & $0.01$ &  		& \multicolumn{1}{c|}{$0.02$} 	&$0.02$	 & $0.02$\\ \cline{1-4} \cline{6-8} 
			\mycaset{10}&1 & $<0.01$ &\myfailh \\
			&2 & $<0.01$ &		\\
			\bottomrule \\
			\multicolumn{8}{l}{\makecell[l]{\fno: In cases 6 and 7, the error only appears in layer 2.}} \\
			\multicolumn{8}{l}{\makecell[l]{\fnt: In Case~10, the effect of the structural error, i.e., missing hyperdiffusion, is too small in short-term\\ and does not show up in analysis increment.}}\\
			\multicolumn{8}{l}{\makecell[l]{\fnth: The failure of ANNs (as interpolator) is due to spectral bias discussed in \cref{sec:neuralfitobs}. \\ This can be addressed by   adding an RFF layer (ANN+RFF), extending \method\ to sparse observations.}} \\
		\end{tabular}
	}
\end{table}

\def\thirdwidth{0.4}
\def\myscale{0.25}
\def\myscaleO{1.22}
\def\myscaleT{1.15}
\def\mycrop{15}
\def\mycropO{15}
\begin{figure}[!ht]
\def\casenum{3} \def\layernum{1}
\begin{subfigure}[b]{\thirdwidth\textwidth}
	\caption{Case \casenum, Layer \layernum}
\includegraphics[trim=0 0 0 \mycropO,clip,scale=\myscaleO]
{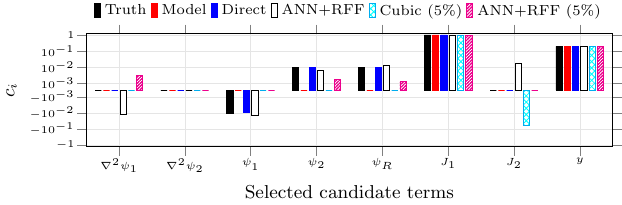}
\label{fig:c3l1}
\end{subfigure}
\\
\def\casenum{5} \def\layernum{2}
\begin{subfigure}[b]{\thirdwidth\textwidth}
\caption{Case \casenum, Layer \layernum}
\includegraphics[trim=0 0 0 \mycropO,clip,scale=\myscaleO]
{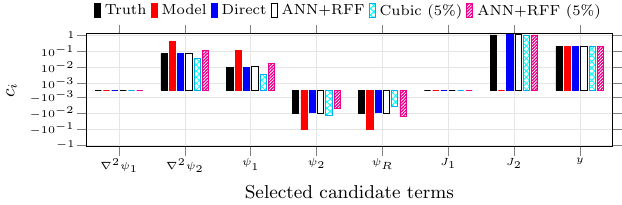}
\label{fig:c5l2}
\end{subfigure}
\\
\def\casenum{8} \def\layernum{1}
\begin{subfigure}[b]{\thirdwidth\textwidth}
\caption{Case \casenum, Layer \layernum}
\includegraphics[trim=0 0 0 \mycropO,clip,scale=\myscaleO]
{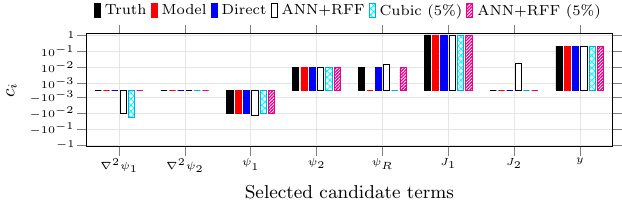}
\label{fig:c8l1}
\end{subfigure}
\caption{
The true, model's, and corrected model's coefficients for a few selected library terms and three representative cases: a) Case 3 (layer 1), b) Case 5 (layer 2), and c) Case 8 (layer 1).
The cases are defined in~\cref{tab:qg_cases} and discovered terms correspond to the reported results in \cref{tab:qg_result}. 
The discovery is performed on full observations (results are shown for direct and \gls{ann}+RFF), and also $5\%$ sparse observations (using cubic and \gls{ann}+RFF interpolation schemes). The legend is: \includegraphics[trim=45pt 90pt 10pt 0pt, clip,scale=1.2]
{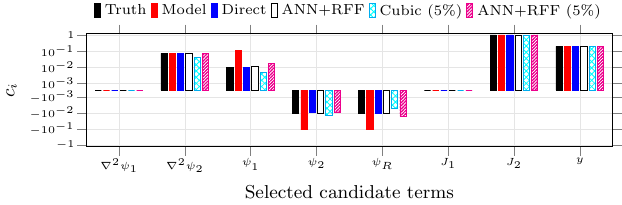}.
The vertical axis is in logarithmic scale showing the model's and corrected model's coefficients for each of the selected library terms, i.e., the prominent terms in~\crefrange{eq:qg}{eq:qg_q}.
While the discovery using direct observations is consistently successful for full observations, the \gls{ann} fit to the same observations consistently fails (results not shown here, discussion in~\cref{sec:neuralfitobs}). The proposed use of \gls{ann}+RFF significantly alleviates this issue by reducing the spectral bias. See the discussion in \cref{sec:neuralfitobs} on the few incorrectly discovered additional terms with ANN+RFF (e.g., $0.01 \times \nabla^2 \psi_1$ and $0.01 \times J_2$ in (\subref{fig:c3l1}) and (\subref{fig:c8l1})).
}
\label{fig:discovery_selected}
\end{figure}

\begin{figure}[!tb]
\def\thirdwidth{0.32}
\centering
\begin{subfigure}[b]{\thirdwidth\textwidth}
\centering
\caption{}
\includegraphics[trim={50 275 550 70},clip,width=\textwidth]{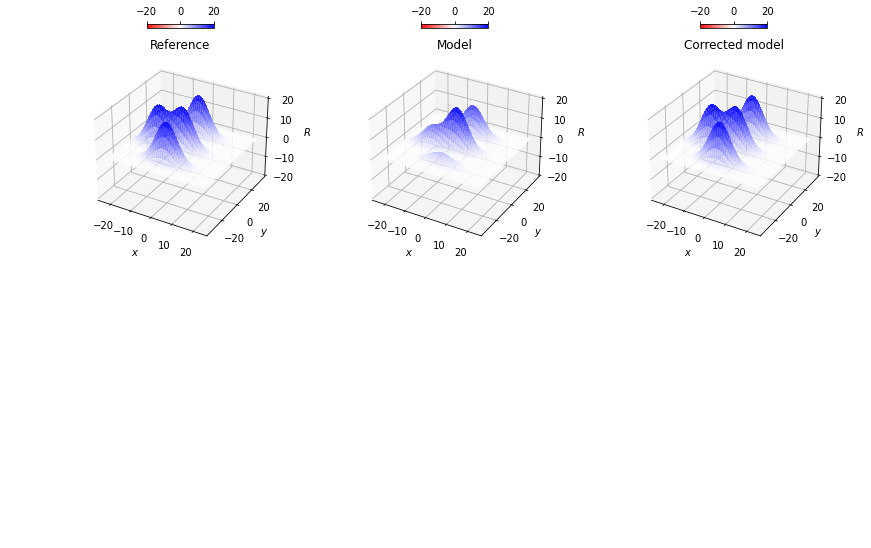}
\label{fig:R_truth}
\end{subfigure}
\begin{subfigure}[b]{\thirdwidth\textwidth}
\centering
\caption{}
\includegraphics[trim={325 275 270 70},clip,width=\textwidth]{mountain_02.png}
\label{fig:R_model}
\end{subfigure}
\begin{subfigure}[b]{\thirdwidth\textwidth}
\centering
\caption{}
\includegraphics[trim={600 275 0 70},clip,width=\textwidth]{mountain_02.png}
\label{fig:R_corrected}
\end{subfigure}
\caption{The orography's profile $R(x,y)$ of the true system (\subref{fig:R_truth}), and
the imperfect model (\subref{fig:R_model}). 
\gls{medida} successfully corrects the profile as shown in~(\subref{fig:R_corrected}).
} 
\label{fig:orography}
\end{figure}

\subsection{Spectral bias in \gls{ann}-based representation of broadly multi-scale states}
\label{sec:neuralfitobs}

The speed and low cost of inference, and the possibility of \gls{ad} and differential modeling have made \glspl{ann} increasingly attractive for applications involving \glspl{pde}. In climate/weather modeling, such applications range from \gls{da}~\cite{Farchi_JAMES_2023} to online-learning of parameterizations~\cite{Frezat_JAMES_2022}, and to ambitious attempts such as discovery of full governing equations~\cite{Chen_NatC_2021, shankar2023differentiable} and more \cite{shen2023differentiable}. Here, we are particularly interested in the interpolation capabilities of \glspl{ann}, as described in \cref{sec:sparse}.
In this section, we evaluate the performance of \glspl{ann} in representing broadly multi-scale states, as a prerequisite for their use as interpolators for applications involving turbulent flows (\cref{sec:missingobs}).
In these experiments, similar to \cref{sec:fullobs}, the state variables are fully observed (without sparsity); however, the model error and the library terms are computed using the output of the trained \glspl{ann}, as further explained below.


For each observed sample at time $t_i$, we train an \gls{ann} {for each state variable} to fit (i.e., represent) the fully observed states (without sparsity), a map between each grid point in the domain and the state variables, $\psi^o_k$ ($k=1$ or $2$). In other words, using the notation introduced in Section~\ref{sec:sparse}, the input of the \gls{ann} is the coordinate (location) of observations, $(x_\ell,y_\ell)$, and the output of the network is the corresponding state variable at that location, $\psi_k^o(x_\ell,y_\ell)$, where $\ell=1, 2, \dots, L$ (no sparsity, $S=0$). Thus, the network is trained to fit/represent the full state. Next, the output of the trained \gls{ann} is used as the observation in \method. One can still use conventional numerical schemes such as spectral or \gls{fd} on the outputs of the \gls{ann} to compute the derivatives and build the library. Alternatively, because this ANN-based representation of the state is ``differentiable'', one can use \gls{ad} to compute derivatives/library. Note that the same steps are followed later when interpolation is needed to deal with sparsity (where $S > 0$). 

Using this approach for fully observed states, we have trained \glspl{ann}, which based on common metrics such as pattern correlation $R^2$ and relative \gls{rmse}, seem to provide very accurate representations of the state variables ($R^2$ around $0.99$, relative \gls{rmse} around $0.25\%$). 
Next, using these trained \glspl{ann}, we have repeated the experiments of \cref{sec:fullobs}, but found the discovery with \method\ to fail in all cases, i.e., $\errorcoeffdiscovered>100\%$ (\cref{tab:qg_result}). This is regardless of which method (AD, spectral, \gls{fd}) is used to compute the derivatives. As a reminder, these observations are not sparse and the direct method of just applying \method\ works very well, as already discussed in Section~\ref{sec:fullobs}. Note that in these experiments, any failure in model error discovery should be due to these seemingly accurate ANN-based representations of the full states.

\begin{figure}[!tb]
\def\thirdwidth{0.32}
\centering
\begin{subfigure}[b]{\thirdwidth\textwidth}
    \centering
    \caption{Classic \gls{ann}}
    \includegraphics[width=\textwidth]{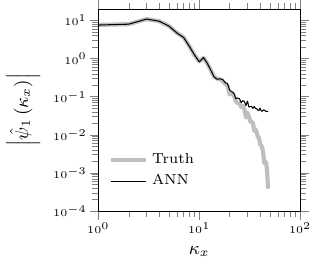}
    \label{fig:psi_norff}
\end{subfigure}
\begin{subfigure}[b]{\thirdwidth\textwidth}
    \centering
    \caption{Classic \gls{ann}}
    \includegraphics[width=\textwidth]{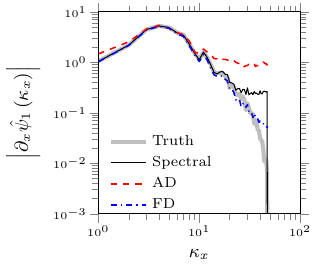}
    \label{fig:psi_dx_norff}
\end{subfigure}
\begin{subfigure}[b]{\thirdwidth\textwidth}
    \centering
    \caption{Classic \gls{ann}}
    \includegraphics[width=\textwidth]{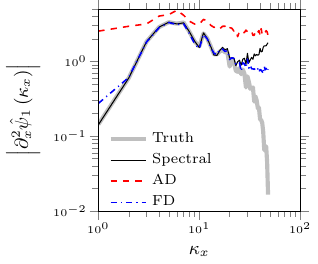}
    \label{fig:psi_dxx_norff}
\end{subfigure}
\\
\begin{subfigure}[b]{\thirdwidth\textwidth}
    \centering
    \caption{\gls{ann}+\gls{rff}}
    \includegraphics[width=\textwidth]{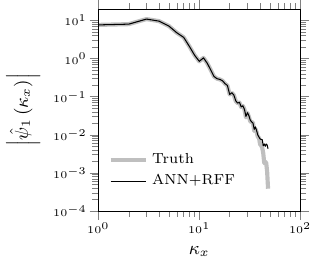}
    \label{fig:psi_rff}
\end{subfigure}
\begin{subfigure}[b]{\thirdwidth\textwidth}
    \centering
    \caption{\gls{ann}+\gls{rff}}
    \includegraphics[width=\textwidth]{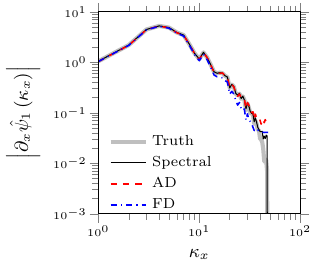}
    \label{fig:psi_dx_rff}
\end{subfigure}
\begin{subfigure}[b]{\thirdwidth\textwidth}
    \centering
    \caption{\gls{ann}+\gls{rff}}
    \includegraphics[width=\textwidth]{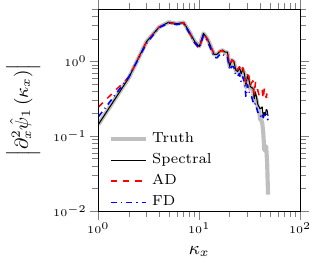}
    \label{fig:psi_dxx_rff}
\end{subfigure}
\caption{Examples of the power spectra of the state variable $\psi_1$  and its first and second spatial derivatives from the numerical solver (truth), and from interpolators that use  (\subref{fig:psi_norff})-(\subref{fig:psi_dxx_norff}) classic \gls{ann}  or  (\subref{fig:psi_rff})-(\subref{fig:psi_dxx_rff}) \gls{ann}+\gls{rff}. Spectra in the zonal direction have been averaged over latitudes. Note that the \gls{ann}-based interpolartors are trained on the full (non-sparse) state variables. \Gls{ad}, \gls{fd} and spectral methods for computing the derivatives are compared in (\subref{fig:psi_dx_norff})-(\subref{fig:psi_dxx_norff}) and (\subref{fig:psi_dx_rff})-(\subref{fig:psi_dxx_rff}).
}
\label{fig:psi_spectrum}
\end{figure}

A detailed analysis shows that the ANN-based representation is only accurate for the large scales and poorly reproduces the small scales (i.e., the high wavenumbers). To demonstrate this, examples of power spectra of the state and its first- and second-order spatial derivatives with respect to the zonal direction are shown in \cref{fig:psi_spectrum}. It is clear that the \gls{ann} has captured the large-scale features of the state accurately (here, for $\kappa_x<15$), which leads to high pattern correlation and small \gls{rmse} ($\kappa_x$ is the wavenumber in zonal direction). However, the spectra clearly deviate from that of the truth at small scales (here, for $\kappa_x>25$). 

The poor representation of small-scales in the state variables significantly degrades the accuracy of derivatives (and thus, the bases in the library). Figure~\ref{fig:psi_spectrum} also compares the accuracy of \gls{ad}, spectral, and second-order central \gls{fd} schemes in computing the first and second derivatives of the output of the trained~\glspl{ann}. The spectral (Fourier) derivative of the full state (from the numerical solver) is considered as the truth. All derivatives are highly inaccurate. Not surprising, the \gls{fd} scheme is least affected by the inaccuracy of high wavenumbers, though it is less accurate at  lower wavenumber compared to the spectral method. Surprisingly, \Gls{ad} is the least accurate method, implying its high sensitivity to the quality of the \gls{ann}-based interpolators.

The loss in accuracy at high wavenumbers is similar to using a coarser effective resolution, here, almost half of the actual resolution.
In other words, the output of a  naively trained \glspl{ann} is only as accurate as that of the state represented on a coarse grid with half of the resolution of the observation. However, there is a subtle but consequential distinction: while representing the state on a coarser grid smoothens the representation, \glspl{ann} inject relatively large errors to high wavenumbers.
Since this error is limited to the small scales, it does not show up in common metrics of accuracy such as pattern correlation and \gls{rmse}.

The origin of this error (the poor representation of small scales) is ``spectral bias'', a well-known phenomenon in the ML literature~\cite{Rahaman_PMLR_2019, JohnXu_arxiv_2022}.
{Although \glspl{ann}, as universal approximators~\cite{Hornik_NN_1989}, can fit any observations (given a large-enough network size and training set), it has been shown that the small-scale features are slower to be learned, to the extent that \glspl{ann} trained with traditional loss and activation functions, regardless of the architecture, are practically incapable of representing the full range of scales. While this inductive bias can be actually useful in many common ML applications (in which the signal is only in the large scales), it can cause major issues for applications involving physical systems, especially those with broadly multi-scale features and no scale separation. One prominent example of such systems is turbulence. For instance, see the paper by~\citeA{chattopadhyay2023long}, which identified spectral bias as the root cause of the long-term instability of data-driven weather models and offered a solution based on a regularized loss function. 


There are various ways to reduce spectral bias \cite{JohnXu_arxiv_2022, chattopadhyay2023long}. In the current application, i.e., using \glspl{ann} as interpolators, an easy and powerful solution is to equip the \gls{ann} with a \gls{rff} layer~\cite{Tancik_neurips_2020}, which has been shown as a promising approach for canonical multi-scale \glspl{pde}~\cite{Sifan_CMAME_2021}. Details of the ANN+RFF approach are presented in \cref{sec:interp}. Briefly,
\gls{rff} is a differentiable but non-trainable layer that slightly and randomly perturbs the \gls{ann}' input, which in our application, is the grid location $(x_\ell,y_\ell)$. The perturbation is done via a nonlinear mapping using sine and cosine functions with random wavenumbers (Eq.~\eqref{eqn:RFF}). Adding the \Gls{rff} layer helps with learning the small-scale features before the network is over-fitted on (biased towards) the large-scale features.

Figure~\ref{fig:psi_spectrum}(\revt{d}) shows that adding the \gls{rff} layer significantly reduces the spectral bias in the interpolator, which now can almost perfectly represent the state across the scales, even at the highest wavenumbers. Consequently, the representations of the derivatives (and this the library) also substantially improve (panels (e)-(f)). Not surprisingly, there are still some differences in the performance of different differentiation schemes, especially for higher-order derivatives. Most notably for \gls{ad} and to some degree \gls{fd}, the noise from the \gls{rff} induces noise at the high wavenumbers, which results in relatively noisy derivatives at high wavenumbers. Nevertheless, the ANN+\revt{\gls{rff}} with any differentiation scheme is much more accurate compared to the classic \glspl{ann}. Overall, the Fourier spectral method shows the best performance, and we use this method hereafter for building the libraries.

Next, we repeat the experiments of Section~\ref{sec:fullobs} but now using the ANN+RFF to represent the full observed states. As shown in \cref{tab:qg_result}, all model errors are successfully discovered, with $\errorcoeffdiscovered<2\%$, which is a significant improvement over when ANN alone was used ($\errorcoeffdiscovered>100\%$), and comparable to the performance of the Direct approach ($\errorcoeffdiscovered<1.3\%$).

}
In \cref{fig:c8l1}, we further examine the performance of ANN+RFF for a few cases where $\varepsilon^*$ is $\order{1}\%$. Moreover, for layer 1 in Cases 3 and 8, \errorcoeffdiscovered is larger than \errorcoeffmodel. Studying layer 2 in Case 5 shows that while $\varepsilon^*=1.45\%$, it is significantly decreased from the original model error of $\varepsilon_m=103\%$. No additional term has been incorrectly identified but a few of the coefficients, e.g. for $\psi_1$ and $\psi_R$ are slightly larger than what they should be. Note that in this example, the \errorcoeffdiscovered is much smaller than \errorcoeffmodel, since the largest contribution to this metric is from the Jacobian term, overshadowing the success of the method in also discovering the relatively smaller but important terms, i.e.,  coefficients of $\psi_1$, $\psi_2$, and $\psi_R$.

Cases 3 and 8 are more interesting, as in these cases, while the original model errors have been corrected, a few additional terms have been mistakenly added to the corrected model. In Case 3, the source of the structural model error, the missing radiative cooling, has been correctly discovered with the right $\tau_{d_1}$ and $\tau_{d_2}$.
However, additional errors, involving $\nabla^2 \psi_1$ and $J_2$, have been also added, leading to an increase in \errorcoeffdiscovered. That said, the incorrectly added terms have small coefficients of $\order{0.01}$. As a result, $0.01J_2$ is not expected to substantially influence the performance of the corrected model given that a $1J_1$ term already exists in the equations of layer 1. Similarly, in layer 1 of Case 8, \method\ has corrected the most important terms, i.e., the coefficients of $\psi_1$, $\psi_2$, and $\psi_R$; however, again, additional terms involving $\nabla^2\psi_1$ and $J_2$ have been identified too, though their coefficients are small ($\order{0.01}$). Overall, this example shows the importance of interpretable model error discovery using ML, as in the end, the users' domain knowledge can be leveraged to understand if a correction is physical or spurious. 

To summarize this section, we have shown that spectral bias can significantly degrade the performance of \glspl{ann} when used as interpolators for equation-discovery techniques. Furthermore, we have shown that adding an \gls{rff} layer to such \glspl{ann} substantially improve their performance and lead to successful model error discovery, at least when tested on fully observed states. Next, in Section~\ref{sec:missingobs}, we will test this approach on sparsely observed states. But before that, it is worthwhile to further discuss spectral bias and its implications for climate science applications. 

\begin{figure}
	\def\thirdwidth{0.49}
	\def\myscale{1.0}
	\centering
	\begin{subfigure}[b]{\thirdwidth\textwidth}
		\centering
		\caption{$\psi_1$}
		\includegraphics[scale=\myscale]{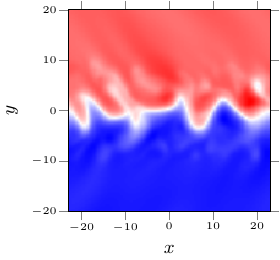}
		\label{fig:psi1_contour}
	\end{subfigure}
	\hfill
	\begin{subfigure}[b]{\thirdwidth\textwidth}
		\centering
		\caption{$\psi_2$}
		\includegraphics[scale=\myscale]{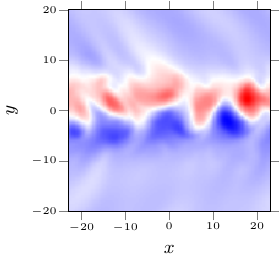}
		\label{fig:psi2_contour}
	\end{subfigure}
	\\
	\centering
	\begin{subfigure}[b]{\thirdwidth\textwidth}
		\centering
		\caption{Image of a fox~\cite{ben_mildenhall}}
		\includegraphics[scale=\myscale]{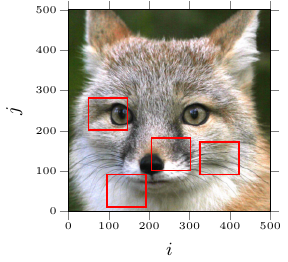}
		\label{fig:fox}
	\end{subfigure}
	\hfill
	\begin{subfigure}[b]{\thirdwidth\textwidth}
		\centering
		\caption{Fourier spectra}
		\includegraphics[scale=\myscale]{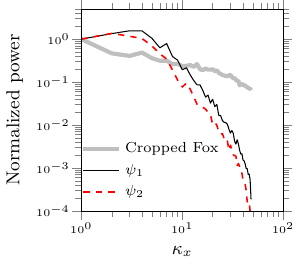}
		\label{fig:spectrum}
	\end{subfigure}
	\caption{
		Spectral contents of arbitrary snapshots of the state variables in QG turbulence (\subref{fig:psi1_contour},\subref{fig:psi2_contour}) compared to a typical natural image~(\subref{fig:fox}).
		The red boxes in (\subref{fig:fox}) are randomly selected and are of the same grid resolution, $96\times84$ as those of the snapshots in (a)-(b). This analysis contrasts the tamed-looking but spectrally complicated (broadly multi-scale) turbulent flows against the complex-looking but spectrally smooth natural images.
	}
	\label{fig:spectrum_all}
\end{figure}

Spectral bias is abundant in ML applications to images and patterns, and is a well-known problem in computer vision, though it often does not lead to major loss of accuracy. However, spectral bias can be severe for many climate processes, such as geophysical turbulence, which involve a very broad range of spatial scales. To show this, we compare the Fourier spectrum of a typical natural image (e.g., a fox) to that of snapshots of the solution of the \gls{qg} equations~(\cref{fig:spectrum_all}). The fox image is cropped to random patches with the same grid size as those of the snapshots of the turbulent flow around the jet region~($96\times84$). The Fourier spectra of these patterns are compared in~\cref{fig:fox}. The magnitude of the Fourier coefficients of the seemingly complex fox image decays $\order{10^1}$ over the entire range of the wavenumbers, {while} the magnitude of the Fourier coefficients of simpler-looking $\psi_{1}$ and $\psi_{2}$ decay over $\order{10^4}$ times. This simple analysis shows the inherently broadly multi-scale nature of turbulent flows (and many other nonlinear physical processes), which can significantly complicate training \glspl{ann} to accurately represent the entire range of scales. That said, not every application in climate science might require an accurate representation of small scales, although small scales often play a role as important as the large scales in the dynamics. Using ANNs as intepolators for equation discovery discussed here is one example where reducing spectral bias becomes essential. The long-term stability of data-driven weather models is another example \cite{chattopadhyay2023long}. There are also a number of recent papers in the climate science literature that have noticed the poor representation of small scales with ANNs for climate processes, although the connections with spectral bias may not have been clear \cite{Rybchuk_arxiv_2023,smith2023temporal,pathak2020using}.




\subsection{Discovery from spatially sparse observations}
\label{sec:missingobs}
Leveraging the success of ANN+RFF as an interpolator, now we repeat the experiments of \cref{tab:qg_result} but with observations that have $5\%$ random sparsity ($S/L=0.05$). ANN+RFF and a baseline bi-cubic interpolation scheme are used following the procedure described in \cref{sec:sparse}. The \gls{ann}-based interpolators are used only to estimate the missing values. The same spectral method is then used on the estimated full state to compute the derivatives and build the library. 

Table~\ref{tab:qg_result} presents the discovery using \method\ with the two interpolators. In all cases, ANN+RFF outperforms the baseline, often by a huge margin. One exception is Case~9, where both interpolators yield comparable, excellent results. \method\ with ANN+RFF successfully discovers the model error in all cases but two and often reduces $\varepsilon$ by a factor of $10$ to $100$. The two exceptions are Case~10, in which as discussed before, the missing hyperdiffusion does not affect the analysis increments, failing the discovery. The other one is Case~3, in which $\varepsilon^*$ remains close to $\varepsilon_m$. Although the structural error (missing radiative cooling) has been discovered and corrected to some degree (by adjusting the coefficients of $\psi_2$, and $\psi_R$), the coefficients are still not that accurate, and an extra term (involving $\nabla^2 \psi_1$) has been mistakenly identified (Fig.~\ref{fig:c3l1}). Still, the performance of the ANN+RFF interpolator is significantly better than the baseline, which cannot identify the structural error at all. Note that other complex cases, such as 5 and 8 show how well \method\ with ANN+RFF works in correcting the model error without adding any additional term (see Figs.~\ref{fig:c5l2} and \ref{fig:c8l1}). 

Overall, \method\ with ANN+RFF successfully discovers model errors from sparse analysis increments, although cases like 3 show the challenging nature of sparse observations. In fact, while $5\%$ sparsity might seem low, state re-construction (super-resolution) from sparse observations of turbulent flows is known to be a challenging task in various areas of science and engineering~\cite{Willcox_CF_2006, Agarwal_EF_2021, Kelshaw_arxiv_2022, Fukami_NMI_2021, Karnakov_arxiv_2023}. Equation discovery from such sparse observations is even more challenging. Here, unlike other studies, we use interpolators trained on only one sample of sparse observations and show promising results once spectral bias is addressed. This progress will enable future studies to focus on developing better interpolators (using several observations and deeper/better architectures) that could handle higher sparsity. For example, \citeA{Fukami_NMI_2021} used deep convolutional neural nets and a sequence of observations to successfully perform super-resolution of turbulent flows with $20\%$ sparsity, although whether the accuracy of the interpolation is enough for higher-derivatives and equation discovery remains to be investigated.

\section{Discussion and Future Work}
\label{sec:Conclusions}

In this \doc, we scaled our recently proposed framework, \method\ \cite{Mojgani_Chaos_2022}, to a much more complex and climate-relevant test case and to spatially sparse observations. \method\ discovers, from analysis increments, close-form equations that represent structural/parametric model errors. As a result, unlike approaches to learning model errors using neural networks, \method\ is data efficient and yields physically interpretable corrections (nudging terms). The interpretability of discovered model errors also increases the likelihood that the corrected models can generalize beyond the training climate, which is a major limitation of deep learning-based model error discovery approaches. Thus, these models corrected using short-term evaluations (at weather timescales) might also lead to more accurate long-term climate projections.  


The numerical experiments explored in this \doc\ show the possibility of discovering different forms of linear and nonlinear structural (and parametric) model errors that are due to various physical processes from analysis increments in a fully turbulent flow. However, we also find an exception, which is while not surprising, has been ignored in most past studies: There are structural errors whose effects do not dominate short-term evaluations (thus the analysis increments), but they have major effects on the long-term evolution of the system. Here, we show such an example based on missing hyperdiffusion, but there are other physical processes, such as cloud microphysics, that can behave in this way as well. While a deeper examination of the analysis increments might provide some information about the effects of such processes, it will be hard to separate their effects from those of numerical errors, at least for the purpose of interpretable model error discovery. 

Overall, we find that if the state of the truth is available without any noise or sparsity, and if the model error dominates the analysis increment, then \method\ can robustly discover complex model errors. These promising results already suggest that \method\ can be a potentially useful method for various model-error-learning applications, because in many of such applications, as explored in recent studies but using deep neural networks~\cite{Wattmeyer_GRL_2021,Bretherton_2022_JAMES, clark2022correcting,Gregory_arxiv_2023,Bora_arxiv_2023} noise-free, full observations of the ``truth'' are available from reanalysis or high-resolution simulations. 

Note that while in this paper we focus on model errors that are missing or incorrect terms in the governing equations, a dominant source of model error in climate science is the lack of numerical resolution, requiring the use of subgrid-scale parameterizations in low-resolution models. Such parameterizations can be learned from data, e.g., from filtered and coarse-grained high-resolution simulations \cite{Zanna_GRL_2020, yuval2020stable, sun2023quantifying}. However, applying equation-discovery techniques to such filtered, coarse-grained data has shown, robustly, the emergence of the nonlinear gradient closure model, which is known to lead to instabilities when coupled to the low-resolution models \cite{Zanna_GRL_2020, Jakhar_arxiv_2023}. One interesting future direction is to apply \method\ between the low-resolution simulations (model) and high-resolution simulations (truth, system) and discover closed-form parameterizations from the analysis increment.  
\revt{Note that a few studies that use differentiable climate models \cite{Frezat_JAMES_2022, Kochkov_arxiv_2023} have recently trained neural network-based parameterizations in a similar “online” fashion: to correct the discrepancy between the trajectories of low-resolution and high-resolution simulations. However, such neural network-based parameterizations, while powerful and already scaled to \glspl{gcm}, remain uninterpretable and will likely not generalize out-of-distribution. These shortcomings can be addressed by approaches such as \method, though scalability to \glspl{gcm} needs further exploration.}


In this paper, successful model error discovery from sparse observations is made possible through the use of \gls{ann}-based interpolators, but only after a major shortcoming of such interpolators is addressed. In general, the use of \glspl{ann} for \glspl{pde} is implicitly based on the assumption that well-trained \glspl{ann} can accurately represent the states as well as their relevant high-order derivatives. 
Here, we demonstrate that spectral bias in \glspl{ann} is a fundamental hindrance to accurate representation of states and more importantly, higher-order derivatives, when applied to turbulent flows. While \glspl{ann} are highly expressive and can fit to any arbitrary input-output map~\cite{Hornik_NN_1989}, they also tend to be biased towards learning large-scales (low-wavenumber) features of the data, a phenomenon known as spectral bias~\cite{Rahaman_PMLR_2019,JohnXu_arxiv_2022}. Given that turbulent flows (and some other processes in the climate system) are inherently broadly multi-scale and span a wide range of wavenumbers, a naive application of \glspl{ann} to such systems will lead to poor representations of the high wavenumbers. Such small-scale features do not measurably contribute to the most commonly used loss functions, e.g., $\normL{2}$-norms, and therefore are often neglected in most of the commonly used metrics when the output of a network is compared with the truth. As discussed in the paper, spectral bias may not cause major issues in many applications, but here, we show that it is a major shortcoming of \gls{ann}-based interpolators for equation discovery (see \citeA{chattopadhyay2023long} for an examples of the implications of spectral bias on long-term instability of data-driven weather models).


{To reduce the spectral bias and develop accurate \gls{ann}-based interpolators, we have demonstrated the effectiveness of adding an \gls{rff} layer to the class \gls{ann} architecture. We show that ANN+RFF captures the entire range of scales in turbulent flows, enabling \method\ to successfully discover model errors from sparse observations.

In dealing with real observations (from remote sensing, in-situ), aside from spatial sparsity, another major challenge is noise (measurement errors). Previously, in a simple chaotic test case, we have demonstrated that noise can be adequately reduced using \gls{da} techniques such as ensemble Kalman filters with a large-enough ensemble size for the successful discovery of model errors with \method~\cite{Mojgani_Chaos_2022}. However, the rate of reduction of noise (as a function of ensemble size) in higher-order derivatives of state in large-scale problems can become a practical challenge. One solution to this challenge is using efficient surrogates, such as \glspl{rom} or \glspl{ann} with accurate short-term prediction horizons~\cite{Chattopadhyay_JCP_2023}, to increase the number of ensemble members at a low computational cost.

Interpretable model error discovery from noisy, sparse observations is challenging. While the results presented here are promising in this QG test case and for low-level random sparsity, more extensive explorations using more challenging test cases, e.g., an intermediate-complexity \glspl{gcm} and more realistic synthetic observations, are needed. The spectral bias-reduced ANN+RFF interpolators offer a step toward success in this direction.     


\section*{Open Research Section}
All data are generated running the solver that is made available at {\mydepository}. The codes to evolve the model, generate the library of bases, and perform the discovery, as well as to train the neural networks and interpolation schemes are made available at {\mydepository}.

\acknowledgments
We would like to thank John Harlim and Ebrahim Nabizadeh for insightful discussions. We are grateful to Hamid Pahlavan and Qiang Sun for helpful comments on an earlier draft of this manuscript. This work was supported by an award from the ONR Young Investigator Program (No.~N00014-20-1-2722), a grant from the NSF CSSI Program (No.~OAC-2005123), and by the generosity of Eric and Wendy Schmidt by recommendation of the Schmidt Futures program. Computational resources were provided by NSF XSEDE/ACCESS (Allocations ATM170020 and PHY220125) and NCAR’s CISL (Allocations URIC0004 and URIC0009).
\revt{We also would like to thank the anonymous reviewers for their constructive comments.}

\appendix

\section{ANN with \Gls{rff}}
\label{sec:interp}

To reduce the spectral bias in training of \glspl{ann}, we employ a non--trainable \gls{rff} layer~\cite{Tancik_neurips_2020}. While ANN+RFF has been used to solve canonical \glspl{pde}, for example using \glspl{pinn}~\cite{Sifan_CMAME_2021}, here, for the first time, we demonstrate its effectiveness for equation-discovery applications and the more challenging case of turbulent flows, which are broadly multi-scale.


The \gls{rff} layer alleviates the spectral bias by slightly and randomly perturbing the coordinate-based input. The layer is constructed as (Fig.~\ref{fig:mlprff})
\begin{equation} \label{eqn:RFF}
	\gamma \left(\inputvec\right)=
	\transpose{
		\left[ 
		\cos \left(2 \pi {\mathbf{B}} \inputvec\right),
		\sin \left(2 \pi {\mathbf{B}} \inputvec\right) 
		\right]
	},
\end{equation}
where $\inputvec=(x_\ell,y_\ell)$ is the input to the layer, and each entry in the (constant) noise matrix $\mathbf{B}$ is randomly sampled from a normal distribution with a standard deviation of $\sigma$, a hyperparameter.
The output of the layer is then passed through the classic architecture of an \gls{ann} (\cref{fig:mlprff}). 
The layer directs the output of the network towards learning of high wavenumber by perturbing the eigenvalues of the \gls{ntk} towards a well-behaved eigen-decomposition~\cite{Tancik_neurips_2020}. This addition to the network leads to learning small-scale features before the network is over-fitted (biased) towards the large scale features.

In this \doc, the ANNs are $5$-layer deep, with $1000$ nodes per layer. 
Adam optimizer~\cite{Kingma_arxiv_2014} with learning rates of $10^{-3}$ and $5\times10^{-4}$ are used in all the experiments. We have found $\sigma=1$ to provide the lowest training error in terms of how closely the spectra of the ANN output and truth match.

\begin{figure}
	\centering
	\includegraphics[scale=1.35]{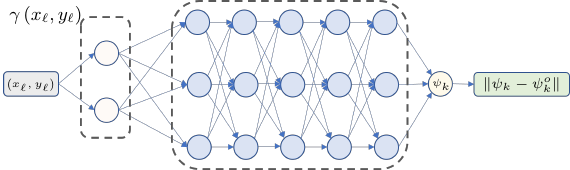}
	\caption{
		An \gls{ann} architecture with a non-trainable \gls{rff} layer to reduce the spectral bias in the interpolator. In this \doc, separate \glspl{ann} are trained for each layer of the \gls{qg} system, i.e., $k=\left\{1,2\right\}$.
	}
	\label{fig:mlprff}
\end{figure}

\bibliography{ModelError.bib}

\end{document}